\documentclass[11pt,a4paper]{article}

\usepackage[top=35mm,bottom=35mm,right=25mm,left=25mm]{geometry}
\usepackage{graphicx}
\usepackage[nolist]{acronym}
\usepackage{units}
\usepackage{algorithmic}
\usepackage{hyperref}
\hypersetup{pdfborder=0 0 0}
\usepackage{floatrow}
\usepackage{subcaption}
\usepackage{amssymb,amsmath}
\usepackage{units}
\usepackage[utf8]{inputenc}
\usepackage{tikz}
\usetikzlibrary{arrows,shapes,calc}
\usepackage{pgfplots}
\pgfplotsset{compat=newest}
\newlength\figH
\newlength\figW
\begin{document}
\title{Multi-Sensor Data Fusion for Accurate Traffic Speed and Travel Time Reconstruction}
\author{Lisa~Kessler$^1$, Felix~Rempe$^2$, and~Klaus~Bogenberger$^1$
	\thanks{$^1$L. Kessler and K. Bogenberger are with Technical University of Munich, Chair of Traffic Engineering and Control, Arcisstrasse 21, 80333 Munich, Germany, e-mail: lisa.kessler@tum.de }%
    \thanks{$^2$F. Rempe is with BMW Group, Mobility Technologies, Petuelring 130, 80788 Munich, Germany}%
}

\maketitle

\begin{abstract}
This paper studies the joint reconstruction of traffic speeds and travel times by fusing sparse sensor data. Raw speed data from inductive loop detectors and floating cars as well as travel time measurements are combined using different fusion techniques. A novel fusion approach is developed which extends existing speed reconstruction methods to integrate low-resolution travel time data. Several state-of-the-art methods and the novel approach are evaluated on their performance in reconstructing traffic speeds and travel times using various combinations of sensor data. Algorithms and sensor setups are evaluated with real loop detector, floating car and Bluetooth data collected during severe congestion on German freeway A9. Two main aspects are examined: (i) which algorithm provides the most accurate result depending on the used data and (ii) which type of sensor and which combination of sensors yields higher estimation accuracies. Results show that, overall, the novel approach applied to a combination of floating-car data and loop data provides the best speed and travel time accuracy. Furthermore, a fusion of sources improves the reconstruction quality in many, but not all cases. In particular, Bluetooth data only provide a benefit for reconstruction purposes if integrated distinctively.

\end{abstract}

\begin{acronym}[GASM]
	\acro{ASM}{Adaptive Smoothing Method}
	\acro{PSM}{Phase-Based Smoothing Method}
	\acro{FCD}{Floating-Car Data}
	\acro{BT}{Bluetooth}
	\acro{IMAE}{Inverse Mean Average Error}
	\acro{ITS}{Intelligent Transportation Systems}
	\acro{MAPE}{Mean Absolute Percentage Error}
	\acro{GNSS}{Global Navigation Satellite Systems}
	\acro{PDF}{Probability Density Function}
	\acro{WMJ}{Wide Moving Jam}
\end{acronym}

\section{Introduction}
For various applications in traffic engineering, it is fundamental to know about the traffic conditions on a road stretch with high certainty and sufficient spatio-temporal accuracy. A complete representation of traffic conditions is especially crucial for understanding traffic flow, for the effectivity analysis of control measures and for training data-driven prediction models. In contrast to real-time or predictive state estimation, these applications are usually applied retrospectively.

The retrospective analysis often focuses on average vehicle speeds per time and space interval on a road since this provides benefits such as enabling the deduction of travel times for road users, providing jam tail warnings \cite{Rempe.2017.MTITS} aiming at the reduction of rear-end collisions at jam tails, etc. However, using current sensor technology, average vehicle speeds are not measured for all times and places on a road stretch. Rather, various types of sensors are available that provide traffic-related data at different times for different places. Raw sensor data must therefore be processed in order to determine an accurate reconstruction of traffic conditions. 

Nowadays, several sensor technologies are in place that gather data, each coming with advantages and disadvantages when applied. Induction loops, that are buried in the road surface, provide very exact and reliable speed information but are mainly limited to few road stretches since the installation and maintenance costs are high. \ac{FCD}, also called probe data, are gathered from vehicles or smartphones that determine their position via \ac{GNSS} and report this position on a regular basis to a central server. Time and space differences allow for reconstructing the probe's speed profile on a road. \ac{FCD} are available wherever traffic is flowing, but represent only a sub-sample of the whole fleet. With WiFi/\ac{BT} sensor technology, the unique MAC address of a device that passes two neighboring stations is registered, allowing the derivation of the travel time and therefore the average speed of devices that pass two neighboring stations~\cite{Martchouk.2011, Margreiter.2016.1, Haghani.2010, Barcel2010TravelTF, Lesani2016ArterialTM}. \ac{BT} installation is not expensive but -- like \ac{FCD} -- the receivers do not collect information from all vehicles and additionally, since they are conceivably placed several kilometers apart from each other, the average speed can be less granular.

Measuring traffic conditions with various sensors offers a great opportunity to increase the accuracy of traffic state estimates. However, the mentioned differences and characteristics of each technology challenge the fusion of the sources. The aim of an advanced fusion method is to make use of all information hidden in the data and compute a combined result that outperforms estimates based on a single source. Additionally, a combination of satisfactorily precise sensor combinations which are available at lower costs might be a reasonable compromise for decision makers, so knowing these combinations would be beneficial to them.

Given various sensor technologies, and various algorithms to process collected data, it is difficult to decide, which technology one should adopt and which algorithm one should deploy. This paper seeks to support decision makers, practitioners and researchers in selecting the combination of sensor data and a reconstruction approach that provides the greatest benefit for their specific problem. Since a real-world application requires algorithms to cope with sparse and missing data, this paper studies approaches with high robustness that can be applied directly. Based on real data collected on a German freeway, various algorithms and combinations of sensor technology are evaluated. Results comprise the accuracy of reconstructed space-time speeds as well as the accuracy of deduced travel times. 

The paper is structured as follows. Section \ref{sec:sota} gives a literature review on the comparison of different traffic data detection systems and on information fusion approaches. Section~\ref{sec:data} describes the study site and data that are used to evaluate subsequent approaches. In section~\ref{sec:methods}, existing applicable fusion methods are briefly summarized. Subsection~\ref{sec:PSM} describes the adaption of the \ac{PSM} to consider \ac{BT} data in a distinct way. Section \ref{sec:eval} presents the applied quality metrics and the obtained results applying the methods to varying sensor setups. The conclusion in \ref{sec:conclusion} wraps up the results and provides potential further research directions.

\section{State of the Art}
\label{sec:sota}
Comparisons of different traffic detection technologies have been widely performed in the past. In \cite{klein2020traffic}, a comprehensive summary of available sensors and fusion techniques is given. The authors of \cite{Bachmann.2013.2} compared Bluetooth measurements and loop detector data in the Greater Toronto Area on a stretch of several kilometers. In \cite{Kessler.2018.CTS}, the authors describe an offline comparison between loop detectors and floating cars, determining which is able to detect a traffic incident earlier. In \cite{Cohen.2015}, the authors statistically analyze the differences between loop detectors and floating car data in the area of Lille, France.

Additionally, different fusion techniques have been investigated. El Faouzi and Klein~\cite{Faouzi.2016} give a survey of current data fusion techniques for intelligent transportation systems. In~\cite{klein2019sensor}, they present three widely applied data fusion techniques and describe their relevance to \ac{ITS}: Bayesian inference, Dempster‒Shafer evidential reasoning, and Kalman filtering. In~\cite{Zeng.2008}, an evidence-theory-based data fusion approach for traffic incident detection is described. Data from inductive detectors, camera observation and floating car data are fused on a rather short stretch of a few hundred meters on an urban highway. The authors of~\cite{Corsi.2011} applied data fusion techniques for traffic planning and control in a setting with satellite images, acoustic and GPS data. In~\cite{Zhou.2015}, the authors describe a real-time capable framework for the fusion of loop detector and GPS data. This framework is able to distinguish lane-based traffic states. The authors of~\cite{faouziTravelTime2009} study the fusion of loop data and toll collection data using a Dempster-Shafer approach in order to get an improved travel time estimate. In~\cite{Yuan.2014}, an approach to network-wide traffic state estimation combining loop detector and floating car data is presented. The authors of~\cite{Rempe.2017.MTITS} developed a model to fuse \ac{FCD} and loop detector data to forecast congestion fronts on a freeway. A comparison of two model-based approaches on filtering methods is conducted in~\cite{Trinh.2019}. The results are confirmed using synthetic data from a simulation. Liu et al.~\cite{Liu.2018} describe an extended Kalman filter method for freeway traffic state estimation fusing two data sources: wireless communication records and microwave sensor detections. Another Kalman filter based approach is given in~\cite{Fulari.2015}. In~\cite{He.2016}, the authors discuss a data fusion approach for cellphone probes and fixed sensors, and give a sensitivity analysis on impact factors. The article~\cite{Chang.2016} describes a data fusion for travel time estimation from toll collection stations and stationary vehicle detectors in Taiwan. Rostami et al.~\cite{ShahrbabakiFusion} propose a fusion of loop data and \ac{FCD} at intersections to estimate queue lengths and outflows. In~\cite{AMBUHL2016184} and~\cite{DAKIC2018317}, also a fusion of loop data and \ac{FCD} is described with the goal to approximate the \emph{Macroscopic Fundamental Diagram} of urban networks. Bachmann et al.~\cite{Bachmann.2011, Bachmann.2013.1} compared seven fusion methods for traffic speeds and travel time estimations. One key finding is that a simple convex combination of loop detectors and \ac{BT} measurements is one of the best fusion strategies. However, data stems from micro-simulations which tend to idealize real data. The authors of~\cite{hegyi_fusion} fuse various sensors, loops, \ac{FCD} and camera data using the \ac{ASM} to improve the speed of jam detection and respective control measures. An evaluation is performed using simulated data.

The mentioned studies are mostly limited to the usage of two different sensors, which limits the applicability in many scenarios. Furthermore, they mostly consider only one quality metric that is investigated, e.g. the spatio-temporal speed distribution or travel time. Some of the mentioned studies focus on the estimation of traffic conditions in dense networks, which is a different challenge than the one emphasized in this paper. Furthermore, data are often derived from micro-simulations, which allow for extensive studies but result in data that are usually more homogeneous and less noisy than real data. If studies utilize empirical data, they often focus on a rather short road stretch which gives an insight into only that specific freeway section.

The approach described in this paper is based on empirical data collected via three common sensor technologies: loop detectors, \ac{FCD} and low-frequency travel time data from \ac{BT} devices on a long stretch of a German freeway. The number of data points is large, which allows a detailed study of all combinations of data as well as several algorithms processing the data. Furthermore, this work applies two metrics which provides insights into the accuracy of both reconstructed traffic speeds and reconstructed travel times. The algorithms compared in this study, are state-of-the-art methods such as the \ac{ASM}, the \ac{PSM}, simple averaging methods and an extension of the \ac{PSM}. This extension is a minor, but effective change to the \ac{PSM}, which allows for the integration of low-frequency travel time data in order to achieve higher reconstruction accuracies. 

\section{Notation \& Data} \label{sec:data}
Speed measurements for all sensors are considered on a road stretch with length $X$ and a time period $T$. The data of all detection technologies are represented as spatio-temporally discrete speed values in a uniform grid with step size $\Delta X = 100m$ and $\Delta T = 60s$. Thus, the domain can be represented as a matrix with $n_X$ rows and $n_T$ columns, where an entry (also called \emph{cell} in the following) is referred to as $(i,j)$, where $i = 1, \dots, n_T$ and $j=1,\dots, n_X$. In each cell, the speed value is constant per data source and is denoted as $v_{i,j}$. Given a set $S$ of sensor technologies on the considered road stretch, $V_{s}, s \in S$ with $S=\{FCD,LOOP,BT\}$ denote the speed matrices of \ac{FCD}, loop detectors and \ac{BT} sensors, respectively. 

\ac{FCD} comprise trajectories of vehicles. A trajectory of one vehicle contains all information that a vehicle, equipped with a \ac{GNSS}, collects about its space-time speed. An equipped vehicle samples its current position $x$ at time $t$ with a certain frequency, and thus generates tuples of $(t,x), t \in [0,T], x \in [0,X]$ along the road stretch. Since no further speed information is given, for simplicity, the vehicle's speed between two sampled positions is assumed to be constant. With sampling frequencies, that are in the same order of magnitude as the time discretization of the domain, this basic assumption is sufficient. In order to turn the piece-wise linearly interpolated vehicle position into grid speeds, for each grid cell which is passed by the vehicle, i.e. the vehicle traveled $\Delta x_{i,j}: \Delta x_{i,j} \ge$ \unit[0]{m} and $\Delta t_{i,j} >$ \unit[0]{s} in that cell, a cell-wise speed is computed as $v_{i,j} = \Delta x_{i,j}/ \Delta t_{i,j}$. All cell-wise speeds of all traces are computed, and subsequently, the speeds of all traces are aggregated. If there are multiple speeds for the same cell, the harmonic mean of all assigned speed values is considered. The respective output matrix comprising all speed data from all equipped vehicles is denominated as $V_{FCD}\in \mathbb{R}^{n_X \times n_T}$. 

Speeds measured by loop detectors are given at discrete positions along the road stretch, and with a temporal resolution of one minute. For each loop detector, the measured speeds are assigned to corresponding cells in the grid. The given name is $V_{LOOP}\in \mathbb{R}^{n_X \times n_T}$.

Low-resolution travel times provided by \ac{BT} are interpolated based on the \emph{Bluetooth Interpolation Algorithm}~\cite{Kessler.2019}. This method considers travel times through predefined cells and weights all crossing paths through any cell according to the share of the path inside the cell in order to obtain an averaged speed distribution $V_{BT}\in \mathbb{R}^{n_X \times n_T}$.    

\begin{figure}[htb]
	\vspace{-0.5cm}
	\begin{center}
		\setlength{\figW}{0.3\textwidth}
		\setlength{\figH}{10cm}
		\input{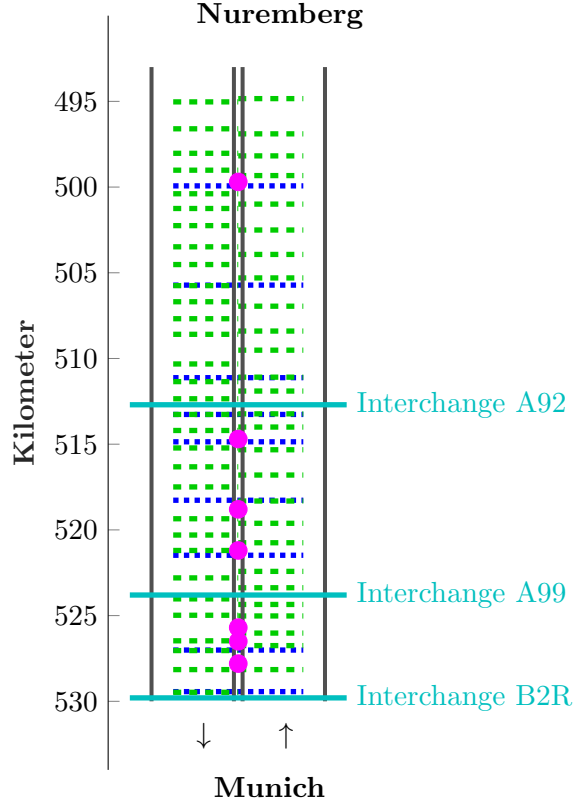}
	\end{center}
	\vspace{-0.5cm}
	\caption{Sketch of considered road stretch: interchanges (cyan), ramps (magenta), induction loops (dashed green), Bluetooth (dotted blue); \ac{FCD} available all throughout the stretch}
	\label{fig:mapOfA9}
\end{figure}

The studies presented in the subsequent section are applied to data collected on May 29, 2019 on German autobahn A9 (Fig.~\ref{fig:mapOfA9}) in the northbound direction during severe traffic congestion. The markers depict the positions of the loop detectors and Bluetooth receivers, respectively. \ac{FCD} is collected from a fleet of cars which are equipped with a \ac{GNSS} device. With sampling times between \unit[5]{s} and \unit[20]{s}, depending on the software version, the vehicle collects positions and timestamps. Packets of positions and timestamps are reported to a central server. In order to ensure privacy, the transmission ID is shuffled from time to time, and some packets are retained such that tracing a vehicle over its entire journey is not possible.

\begin{figure}
	\centering
	\includegraphics[width=0.55\textwidth]{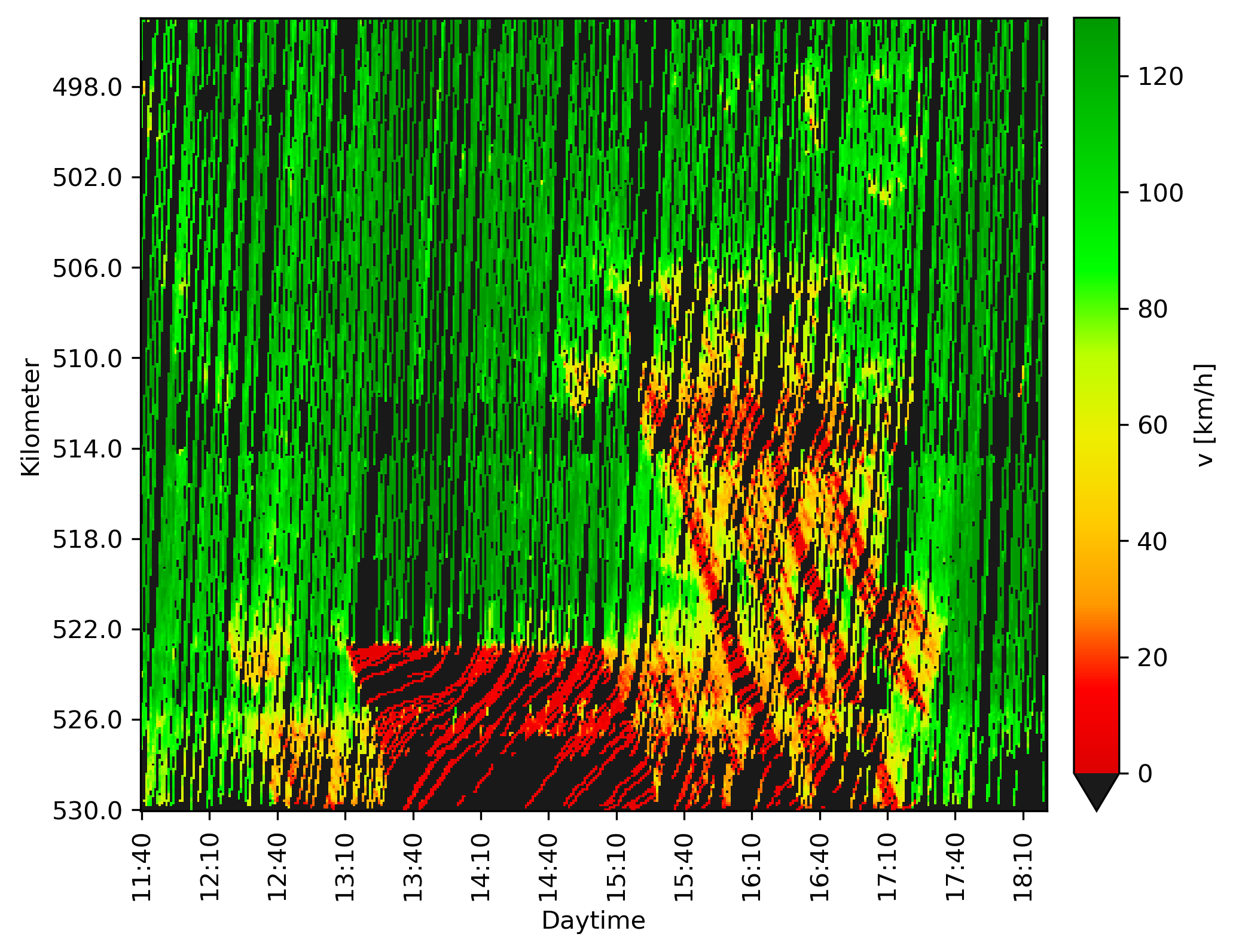}
	\includegraphics[width=0.55\textwidth]{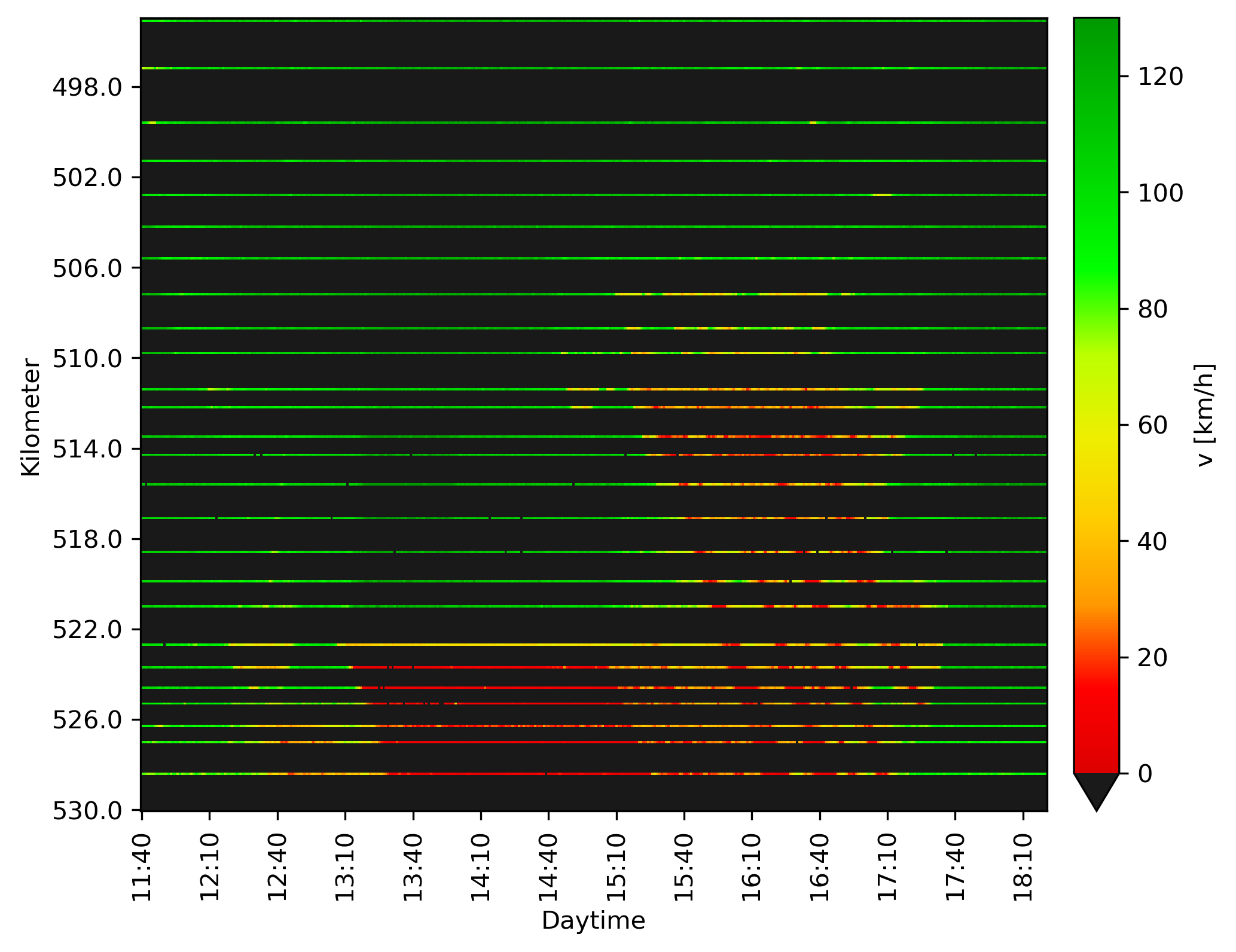}
	\includegraphics[width=0.55\textwidth]{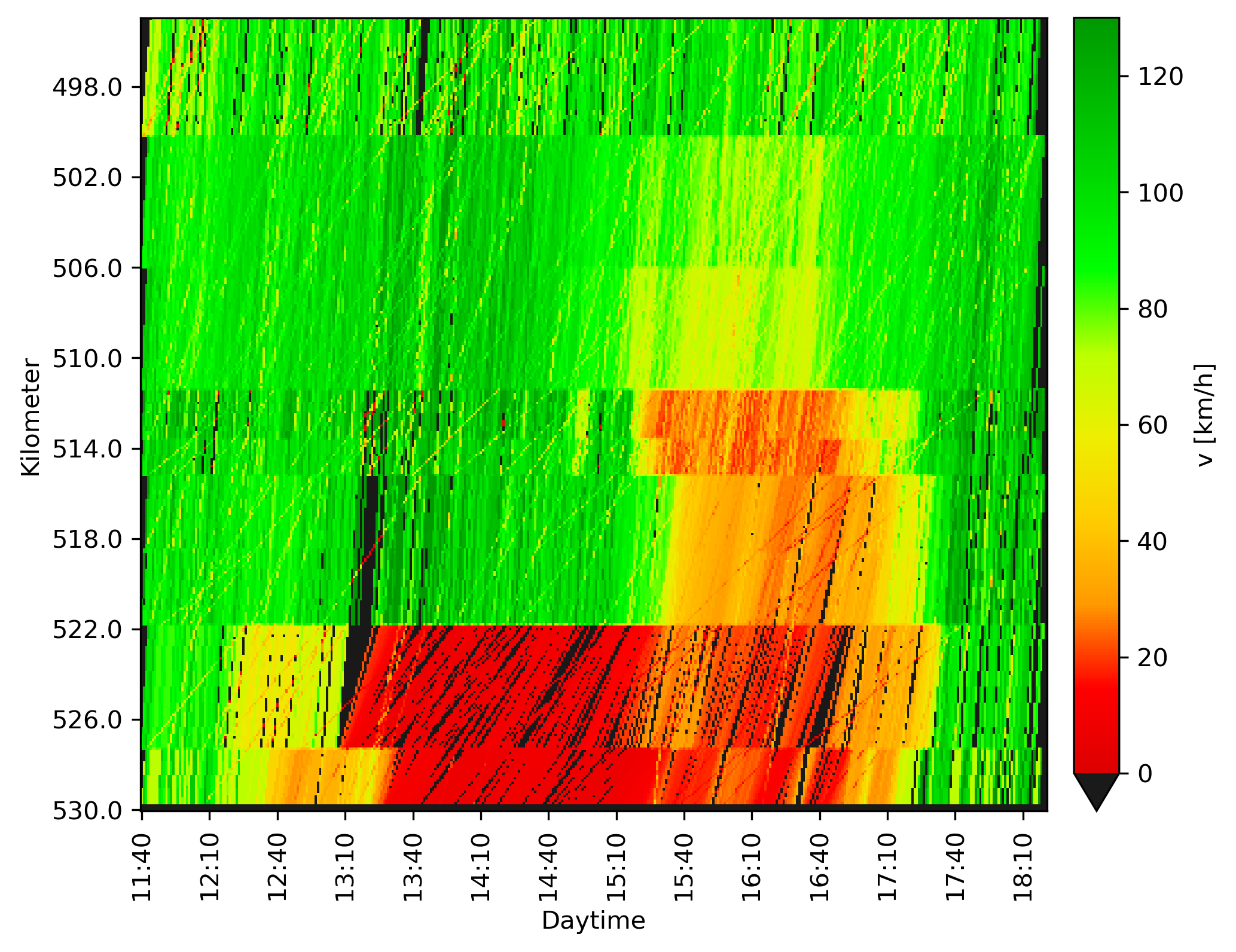}
	\caption{Raw speed data measurements provided by (a) loop detectors, (b) equipped vehicles and (c) Bluetooth receivers}
	\label{fig:v_raw_all}
\end{figure}

All in all, time-discrete data of 27 loop detectors, 11,722 \ac{BT} samples and 1,578 \ac{FCD} traces are available. Fig.~\ref{fig:v_raw_all} displays the raw data. 

\section{Fusion Methods}
\label{sec:methods}
This section presents the fusion methods that are studied in this paper. Three considered state-of-the-art fusion methods are summarized and an extension to the \ac{PSM} is presented.

All methods investigated in the subsequent evaluation require to take as input only gridded speed data. That is necessary, as \ac{FCD} and \ac{BT} contain little information about flow or density. Furthermore, there must not exist requirements regarding minimum data coverage, e.g. a penetration rate of \ac{FCD} or a minimum distance between neighboring detectors. That is necessary to ensure real-world applicability, where a sensor may fail, or where no equipped vehicles may pass a road segment for a longer period of time. Finally, the output of the algorithm must be a continuous speed estimate $V_E \in \mathbb{R}^{n_X \times n_T}$ containing a valid speed $v_{min} \le V_E \le v_{max}\ \forall (i,j)$. 

\subsection{ASM Approach}
\label{sec:ASM}
The \ac{ASM} is a well-known approach used for traffic state reconstruction \cite{Treiber.2002,Treiber.2011,Schreiter.2010,Kessler.2018.TRB} and also for on-line traffic speed estimation \cite{Rempe.2016}. Briefly summarized, raw data of a sparse input source are smoothed in two traffic-charac\-teristic directions: $v_{cong}$ denominating the wave speed in congested traffic conditions, and $v_{free}$ denominating the wave speed in free-flow conditions. In a discrete time-space domain, the resulting complete speed matrices $V_{cong}(t,x)$ and $V_{free}(t,x)$ are combined cell-wise:
\begin{equation} \label{eq:ASM-v}
  V_{ASM}(t,x) = w(t,x)V_{cong}(t,x) + (1-w(t,x))V_{free}(t,x).
\end{equation}
The weight $w(t,x)$ is adaptive and favors low speeds:
\begin{equation} \label{eq:ASM-w}
    w(t,x) = \frac{1}{2} \Big(  1+ \tanh \big( \frac{V_{thr} - \min (V_{cong}(t,x), V_{free}(t,x) )}{\Delta V}  \big) \Big)
\end{equation}
with $V_{thr}$ a threshold where weight $w(t,x)$ equals to $0.5$ and $\Delta V$ a parameter to control the steepness of the weight function. In a theoretical analysis as well as evaluation with real data, van Lint et. al. pointed out that smoothing speeds yields a significant error considering travel time accuracy \cite{Lint.2010}. Instead, they propose smoothing the inverted cell-wise speeds in order to reduce the error. Since travel time accuracy is one of the two key quality metrics in this evaluation, this procedure is adopted in this study, replacing the original formulation of the \ac{ASM}.

Accordingly, for each data source, the discrete space-time matrices $V_S^{ASM} \in \mathbb{R}^{n_X \times n_T}$ are computed. For a fusion, raw data are combined cell-wise and the combined raw data are processed with the \ac{ASM}. In case of at least two data sources providing a speed for the same cell, the harmonic mean is taken. 

\subsection{PSM Approach}
\label{sec:PSM}
The \ac{PSM} is an approach that is based on concepts of the \ac{ASM}. It was developed to reconstruct space-time traffic speeds with higher accuracy given only \ac{FCD} \cite{Rempe.2017.PSM}. It utilizes findings summarized by the \emph{Three-Phase} traffic theory \cite{Kerner.1999,Kerner.2008} in order to distinguish between localized and moving congestion. The method outperformed the \ac{ASM} in a recent study \cite{Rempe.2017.PSM} and is therefore included in the comparison as a state-of-the-art method. We refer to the original paper for a detailed method derivation and evaluation.

Briefly summarized, in the first step of the \ac{PSM}, raw data are smoothed in the direction of typical speed propagation of each traffic phase. $v_{cong}$ is assumed to be the propagation speed of moving congestion with low vehicle speeds (also called \acp{WMJ} in the \textit{Three-Phase} traffic theory). Congestion that is caused by a bottleneck, e.g. a construction site or an on-ramp, is often localized and its downstream front is attached to the bottleneck location. In order to account for the locality, data are smoothed only in temporal direction for the so-called \textit{synchronized} traffic flow phase. Based on the speeds and the amount of available data, each cell $(t,x)$ is classified into one of the three phases: Free flow, synchronized flow or \ac{WMJ} using probability theory.

In the second step, phase-specific speed estimates are computed. Raw speed data that are assigned to a specific phase are smoothed using either a free-flow kernel parameterized with $v_{free}$ or a congested kernel parameterized with $v_{cong}$. The phase-specific speed estimates are aggregated into a final speed estimate using a weighted average. 

The input of the \ac{PSM} are gridded speeds. Additionally, for each cell, a weight matrix $w^{PSM} \in \mathbb{R}^{n_X \times n_T}$ can be given as input to the method. Applying the \ac{PSM} to the raw data of the input sources as well as their cell-wise combinations (see section \ref{sec:ASM}), the respective output matrices $V_S^{PSM}$ are computed. The weight $w^{PSM}$ is set to one for cells with valid data, and zero for cells without data.

\subsection{Extended PSM Approach Considering Low-Frequency Probe Data}
\label{sec:bt_to_v}
In order to apply the mentioned approaches, \ac{BT} data are turned into cell-wise speeds by computing their mean speeds and assigning passed grid cells \cite{Kessler.2019} (see Fig.~\ref{fig:v_raw_all}). However, since the \ac{BT} detectors are usually places several kilometers apart, taking the mean speed of a vehicle is a significant simplification of its real speed. For instance, if there was a mixture of congested and free traffic between two detector locations, a mean speed will smooth all details. For travel time estimations, this approach gives accurate results. In the case, that the space-time speed data is desired, the grid-wise cell speeds lack accuracy. Combining such smoothed speeds with other data sources which deliver more accurate information, will even worsen the resulting output, despite using more data.

Therefore, the idea, presented in this extension, is to introduce a dynamic weight that is assigned to gridded \ac{BT} speed data, which express the trustworthiness of the computed grid speeds. The trustworthiness is influenced by the detector spacing and the measured travel time:

\begin{figure*}[htb]
	\centering
	\includegraphics[width=\textwidth]{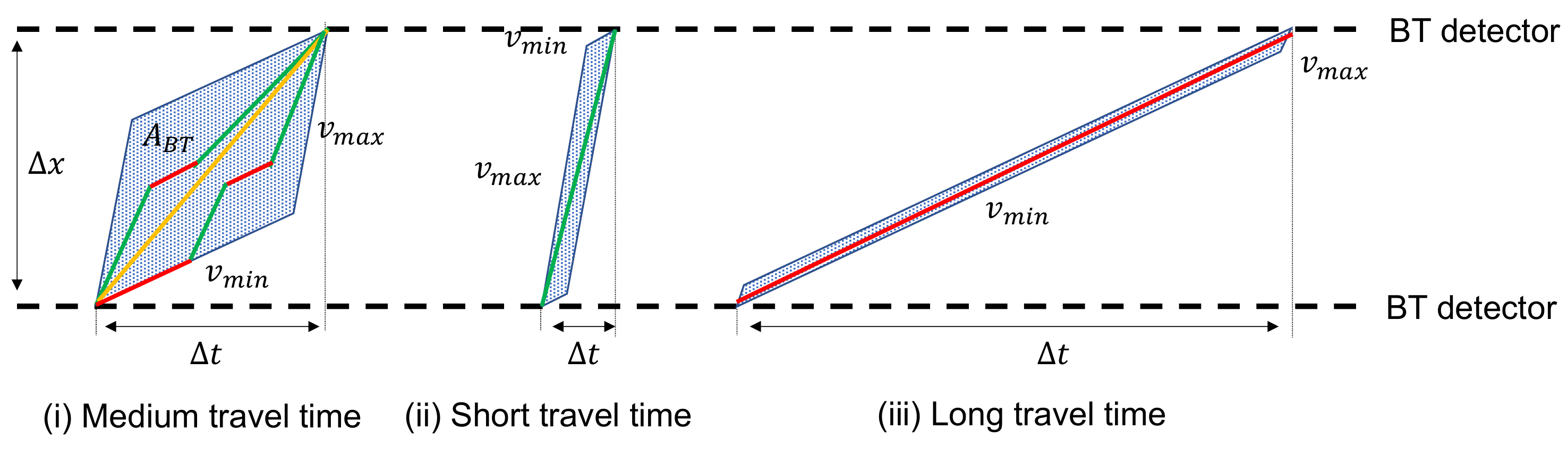}
	\caption{Different travel time measurements and the space of potentially realized trajectories that result in each travel time}
	\label{fig:bt_trace}
\end{figure*}

Assume a vehicle needs time $\Delta t$ to travel distance $\Delta x$ (see Fig.~\ref{fig:bt_trace}). It has a maximum speed of $v_{max}$ and a minimum speed of $v_{min}$. Further assume that the vehicle is not standing, such that $v_{min} > 0$. Then, for an observer who only measured $\Delta t$ and $\Delta x$, it is not known where the vehicle was positioned, and at what speed it was driving, while passing the measured distance in measured time. From the observer's perspective, however, given the assumed minimum and maximum speed, the vehicle's position can be restricted to a certain space-time area. This area is depicted as a parallelogram in Fig.~\ref{fig:bt_trace}, along with three examples of potential vehicle trajectories. Each potential trajectory can be described as a function of the vehicle's position $x(t)$, and its corresponding velocity $v_c(t)$. As illustrated, a medium travel time allows for strong deviations of $v_c(t)$ over time, whereas low travel times restrict $v_c(t)$ to higher speeds. Long travel times can only be realized with vehicle speeds close to $v_{min}$.

A reconstruction method such as the \ac{PSM} is sensitive to wrongly assigned speeds in cells. Therefore, given the chance that the vehicle had a completely different speed profile than the speed profile computed using a simple linear interpolation, the accuracy of the reconstruction suffers. In order to consider the probability of deviation in the reconstruction method, the following approach is implemented:

The variety of potential trajectories is modeled as the space-time area $A_{BT}$ of the parallelogram formed by the time and space difference, and the assumed minimum and maximum vehicle speeds $v_{min}$ and $v_{max}$. The magnitude of the area is supposed to affect the weight of a trace: If the area is large, indicating a great variety of potential trajectories, the weight shall be low. If the area is small, the number of potential trajectories is low and the weight shall be high. An exponential function is utilized to model the decay of the weight $w_{A}$ with increasing $A$

\begin{equation} \label{eq:AtoP}
    w(A) = \exp ( \frac{-A}{\gamma}) 
\end{equation}
with $\gamma \in \mathbb{R}$ a parameter to adjust the sensitivity. The weight $w(A) \in [0,1]$ of all traces passing $(t,x)$ assuming a linear interpolation is averaged and assigned to $w_{BT}$. 

The novel fusion method is denominated as `PSM-W' referring to a dedicated input source weighting of \ac{BT} data. When combining the raw data of loops, \ac{FCD} and \ac{BT}, in this approach the weighted average of all speed cells is taken as input. Raw loop data and \ac{FCD} are assigned a constant weight of one, and said $w_{BT}$ as weight for \ac{BT} data.

\subsection{Section Average}
The `section-average' approach averages collected data in predefined sections. Due to its simplicity, it is still applied in practice and, thus, considered as a relevant approach in this comparison. For each data source, time-space sections are defined and all data that are related to such a section are collected and averaged. Specifically, for loop detectors, section borders are located in the center of two adjacent detector positions. A cell is assigned the speed measurement that is collected by the spatially closest detector at the same moment in time. If, due to an outage of a detector, a measurement is missing, the next closest measurement in time is taken. The resulting speed matrix is denominated as $V^{SEC}_{LOOP}$. 

Start and end times of \ac{BT} samples are collected at the locations of the \ac{BT} detectors. For each section and each time step $\Delta t$, all \ac{BT} traces that cross such a section are identified. The total distance covered by these traces in this section for $\Delta t$ divided by the respective total time of all traces in this section is the resulting average speed at time $t$ for all cells that belong to the section. The resulting speed estimate is denominated as $V^{SEC}_{BT}$. The same approach is done for \ac{FCD}. Compared to stationary detectors, there are no predefined sections. For simplification, the same sections as for detector data are used. In order to assign values to sections without data, a temporal linear interpolation is performed. The resulting matrix is called $V^{SEC}_{FCD}$. Fusions of mutual pairs and all three matrices are simple cell-wise averages of the speeds. 

\section{Evaluation}
\label{sec:eval}

\subsection{Methodology}
\label{sec:methodology}
The aim of an accurate reconstruction method is to generate a complete speed estimate in time and space that is suited to various subsequent applications. Conventionally, the quality of a reconstruction is assessed using speed data only. The drawback is that a potential bias in estimated speeds, e.g. a systematic over-estimation, is not penalized. As a result, estimated travel times over larger segments are erroneous. Therefore, we see it necessary to assess both the accuracy of cell-wise speed estimates and the accuracy of virtual travel times. In the following, the combination of both aspects is considered as the reconstruction quality or accuracy. In order to assess the reconstruction accuracy, the following considerations are taken into account:

\begin{itemize}
	\item[(1)] As visible in the raw data plots, the measurements of each data source are sparse in time and space.
	\item[(2)] Loop detectors provide accurate speed measurements but are limited to certain locations.
	\item[(3)] \ac{FCD} provide relatively accurate speed estimates for varying times and spaces but do not represent macroscopic speeds.
	\item[(4)] \ac{BT}-based travel time measurements are abundant, though the cell-based speeds are inaccurate due to large distances between neighboring stations.
\end{itemize}
For these reasons, in order to assess the space-time speed, those data sources with high spatio-temporal accuracy should be used -- for the evaluation of travel time data, a source with accurate travel time measurements is required. Therefore, a combination of \ac{FCD} and loop detector data assesses the cell-wise speed estimates, and \ac{BT} data are used to assess the travel time accuracy.

\begin{figure}[htb]
	\centering
	\includegraphics[width=0.7\textwidth]{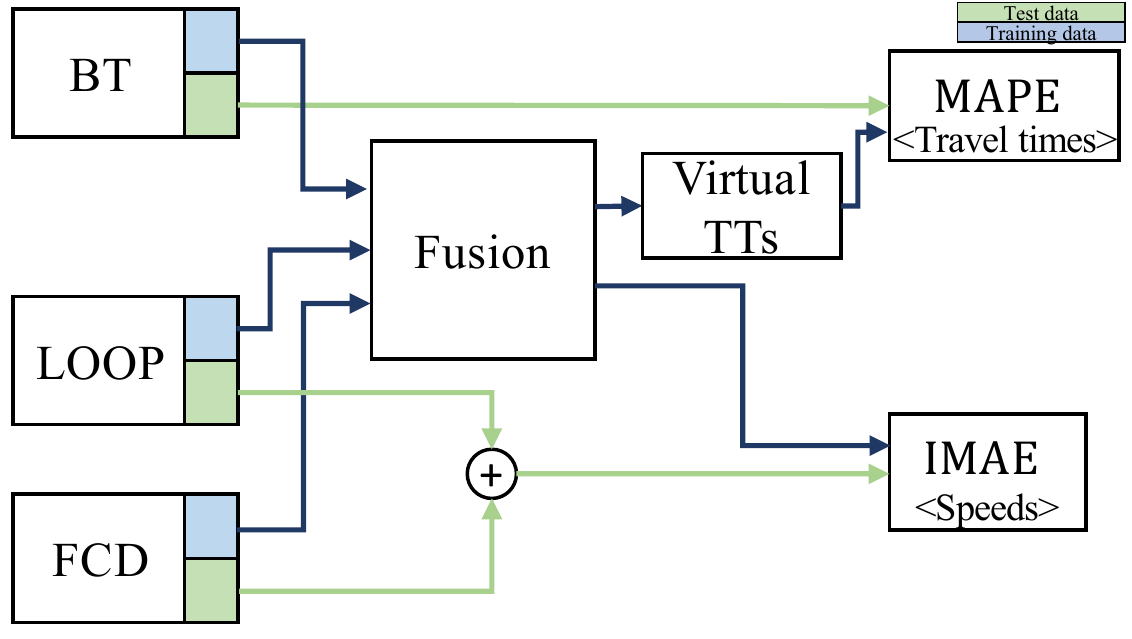}
	\caption{Flow of information of test and training set of sensor data for fusion and quality assessment}
	\label{fig:assessmethod}
\end{figure}

A commonly used approach in model training and evaluation is to divide available data into a training and a test data set. Fig.~\ref{fig:assessmethod} depicts the methodology applied in this evaluation. First, each data source is randomly divided into a training and test set with a ratio of 50:50. Specifically, all speed measurements that are gathered by one detector position are either assigned to training data or test data. \ac{FCD} and \ac{BT} are assigned per trace. Training data are fused in order to generate an estimate $V_E$, and test data of \ac{FCD}, loop detectors and \ac{BT} are used to assess the reconstruction quality.

The quality assessment with a combination of \ac{FCD} and loop data is done using the \ac{IMAE}, eq.~(\ref{eq:imae}). It is a symmetric metric that is sensitive to deviations of lower speeds:
\begin{equation} \label{eq:imae}
	IMAE =\frac{1}{ | v_{test} | } \sum_{v^{i,j} \in v_{test}  } \Big|  \frac{1}{v^{i,j}} - \frac{1}{v^{i,j}_{E}} \Big|
\end{equation}
with $v_{test}$ representing all tuples $v^{i,j}$ that correspond to a cell-wise speed contained in the test set $v_{test}$. The set is defined as the union of all cell-wise speeds in the test sets of \ac{FCD} and loop data.

Quality assessment of travel times with \ac{BT} is based on the comparison of virtual trajectories with the measured traces using \ac{BT} detectors. For each measured trace, a virtual trajectory is computed that starts at the same time and location $(t_{start},x_{start})$ of the real trace. The virtual vehicle drives with the continuous representation of speed $V_E(t,x(t))$ until reaching $x_{end}$:
\begin{equation}
    t_{end} \in [0,T] :  x_{start} + \int_{t_{start}}^{t_{end}}  V_E(t,x(t)) dt = x_{end} 
\end{equation}
Its virtual travel time is defined as:
\begin{equation} \label{eq:vtt}
    VTT=   t_{end}   - t_{start}
\end{equation}
Given $n_{BT}$ as the number of \ac{BT} travel time samples in the test set, $TT_i$ as the measured and $VTT_i, i = 1, ..., n_{BT}$ as the virtual travel times, the \ac{MAPE} is applied as a quality metric. A relative metric reduces the effect of varying segment distances between neighboring \ac{BT} receivers.
\begin{equation} \label{eq:mape}
  MAPE = \frac{1}{ n_{BT} }  \sum_{i=1}^{n_{BT}} \Big|    \frac{VTT_i  - TT_i}{TT_i}       \Big|
\end{equation}
The parameter set for the \ac{ASM} is taken in accordance with \cite{Treiber.2011}. The \ac{PSM} is parameterized according to \cite{Rempe.2017.PSM}. Based on some experiments, $\gamma$ is set to $\unit[500,000]{m\cdot s}$. A formal sensitivity analysis and optimization is left for future work. $v_{min}$ is set to \unit[5]{km/h} and $v_{max}$ is set to \unit[130]{km/h}. The random split between test and training set is done at each run. In total, speed estimation for all scenarios and algorithms as well as quality assessment was done 50 times and average results are presented.

\subsection{Results}
This study intends to give insights into several aspects that come up considering a multi-sensor data fusion. In order to structure the outcomes, the results are examined with respect to two questions:
\begin{enumerate}
	\item Given a certain sensor set-up on a road and several algorithms that can be applied to process raw data, which algorithm returns the most accurate results?
	\item Given the freedom to choose between the three sources of sensor data, which data source or which combination yields best results?
\end{enumerate}

\subsubsection{Algorithm Assessment}
\begin{figure}[htb]
	\centering
	\includegraphics[width=0.65\textwidth]{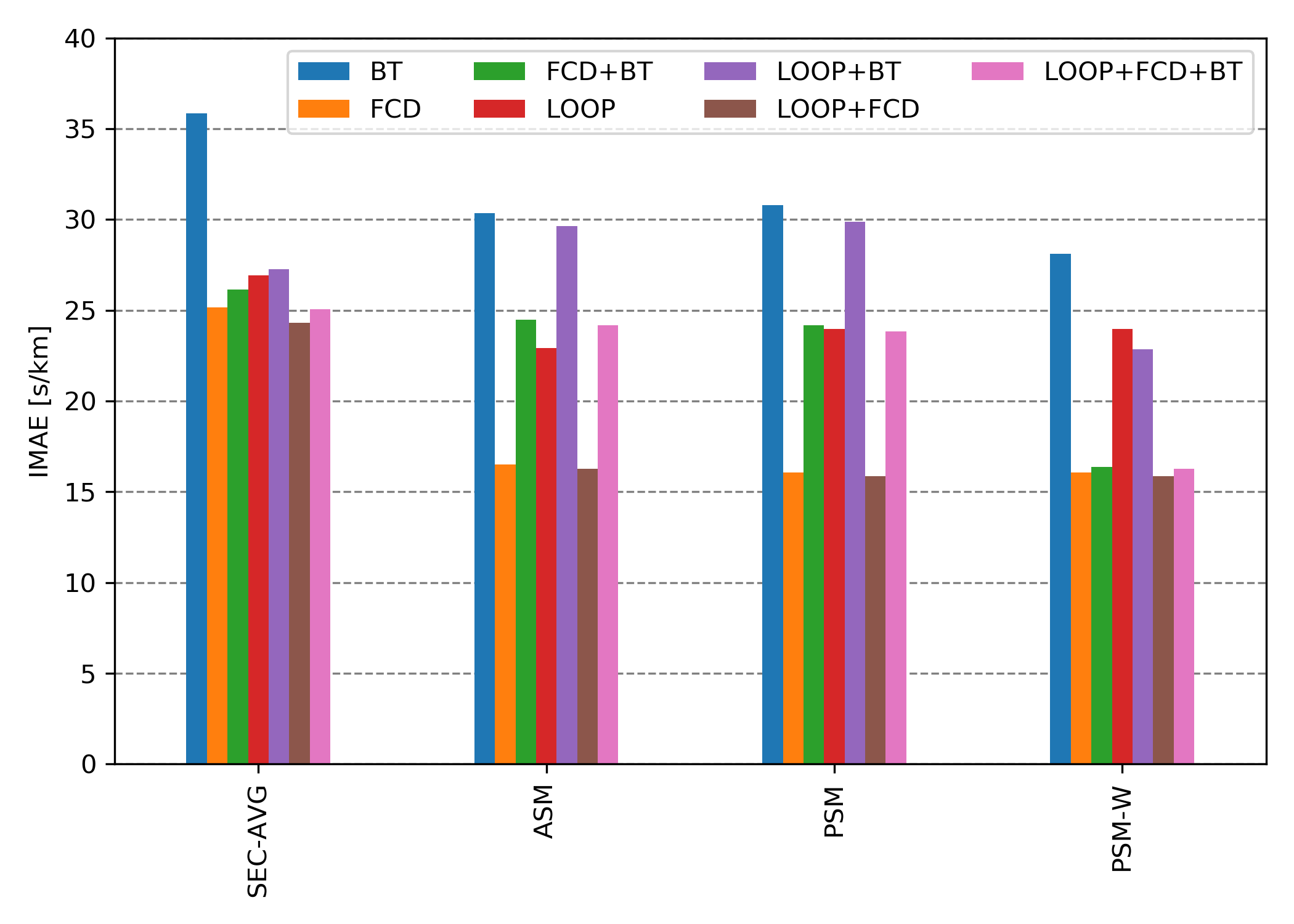}
	\includegraphics[width=0.65\textwidth]{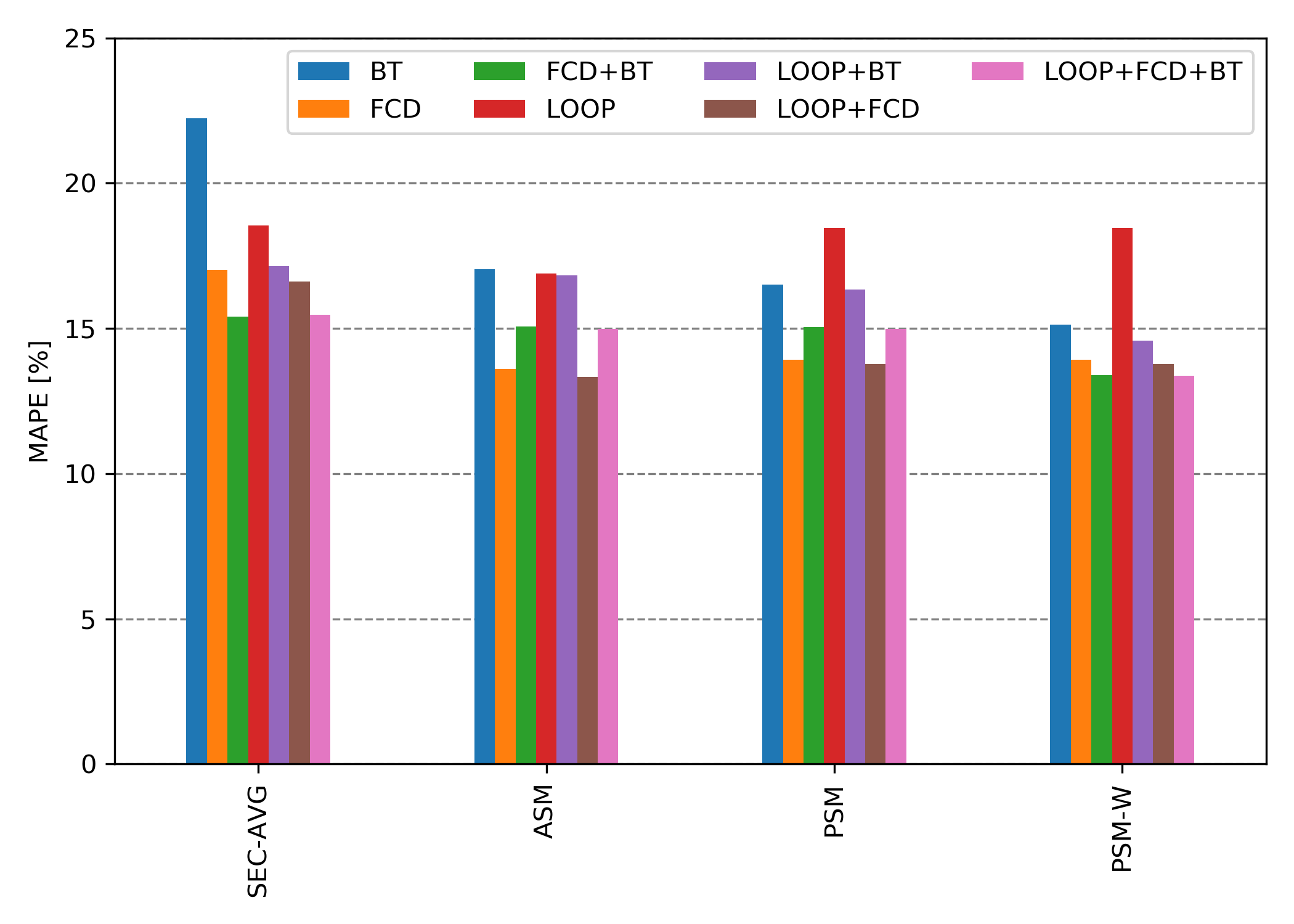}
	\caption{Mean (a) \ac{IMAE} and (b) \ac{MAPE} of all runs with respect to the available sensor technology and the applied algorithm}
	\label{fig:error_vs_algo}
\end{figure}

Fig.~\ref{fig:error_vs_algo} depicts the mean \ac{IMAE} and \ac{MAPE} of all scenarios and algorithms. Several observations can be made: 
\begin{enumerate}
	\item The available sensor data have a significant impact on the resulting errors for each algorithm.
	\item The \ac{IMAE} has a higher variance than the \ac{MAPE}.
	\item Some algorithms perform best with respect to the \ac{IMAE} in a scenario but are outperformed with respect to the \ac{MAPE} (e.g. with \ac{FCD} only, `PSM' has a lower \ac{IMAE} but `ASM' a lower \ac{MAPE}). This shows that both quality metrics measure different properties of an algorithm.
	\item The `SEC-AVG' is the algorithm which results in the lowest accuracy, for \ac{IMAE} as well as \ac{MAPE} in most scenarios. Given only `LOOP+BT', this algorithm has a slight advantage over the `ASM' and `PSM'. Still, the `PSM-W' performs better.
	\item The `PSM-W' performs significantly better in \ac{IMAE} and \ac{MAPE} in all scenarios that involve \ac{BT} data.
	\item On average, the `PSM-W' provides the best quality results. In a `LOOP'-only scenario, the `ASM' performs better.
\end{enumerate}

In order to better understand which estimation errors occurred applying each algorithm, the scenario with all data (`LOOP+FCD+BT') is examined below.

\floatsetup[figure]{style=plain}
\begin{figure*}[p]
    \begin{subfigure}[b]{0.41\textwidth}
    	\centering
    	\includegraphics[width=\textwidth]{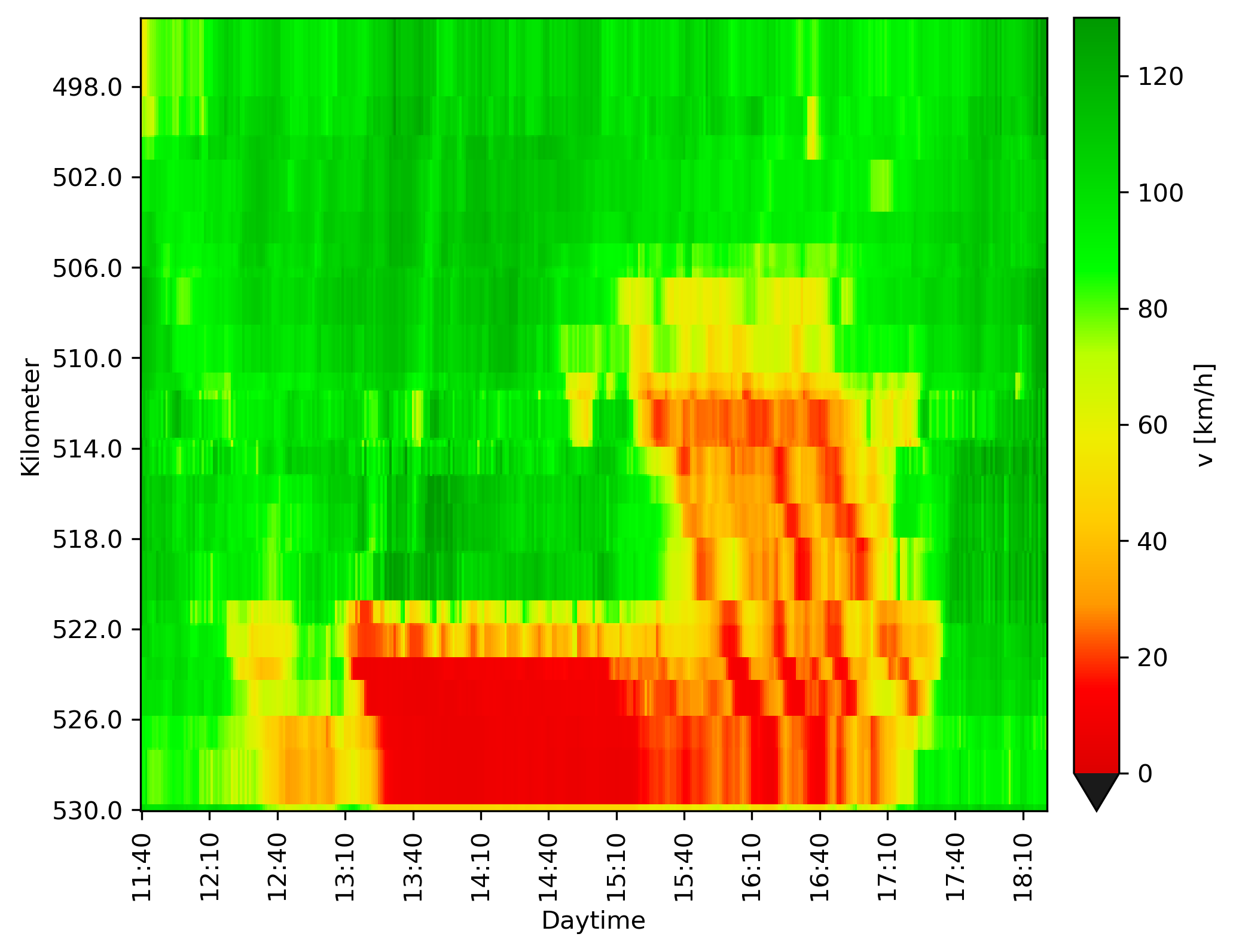}
    \end{subfigure}
    \begin{subfigure}[b]{0.41\textwidth}
    	\centering
    	\includegraphics[width=\textwidth]{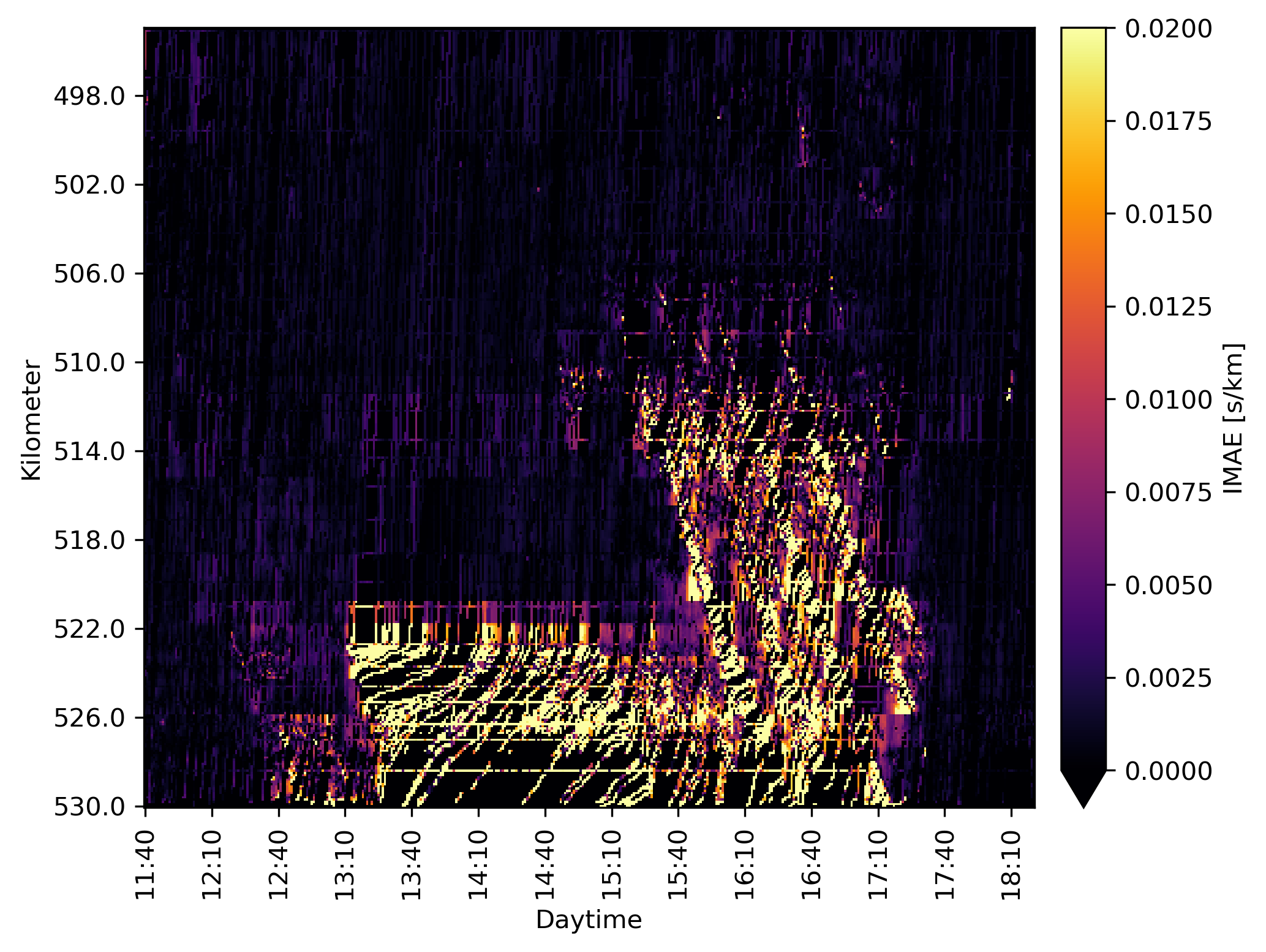}
    \end{subfigure}
    
    \begin{subfigure}[b]{0.41\textwidth}
    	\centering
    	\includegraphics[width=\textwidth]{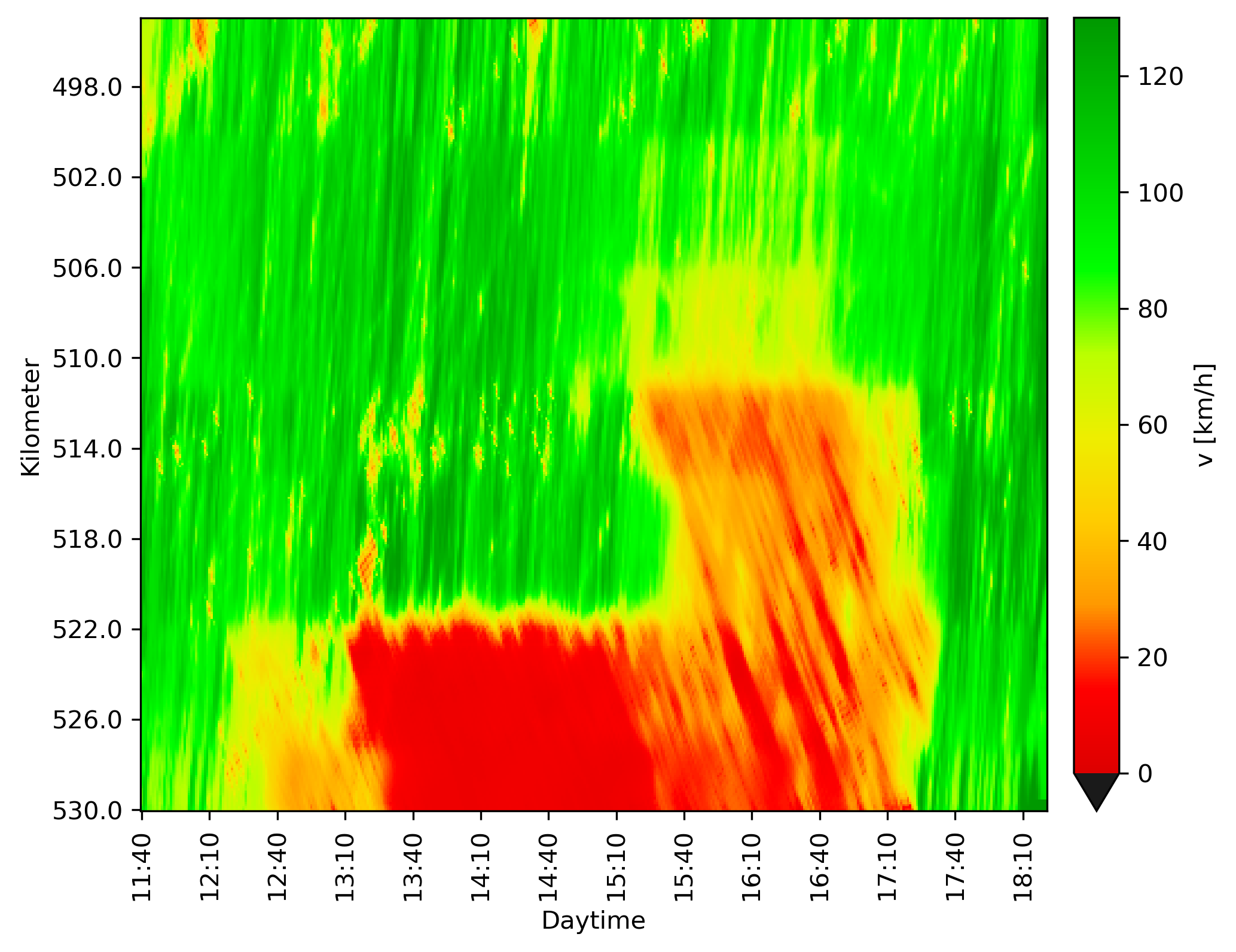}
    \end{subfigure}
    \begin{subfigure}[b]{0.41\textwidth}
    	\centering
    	\includegraphics[width=\textwidth]{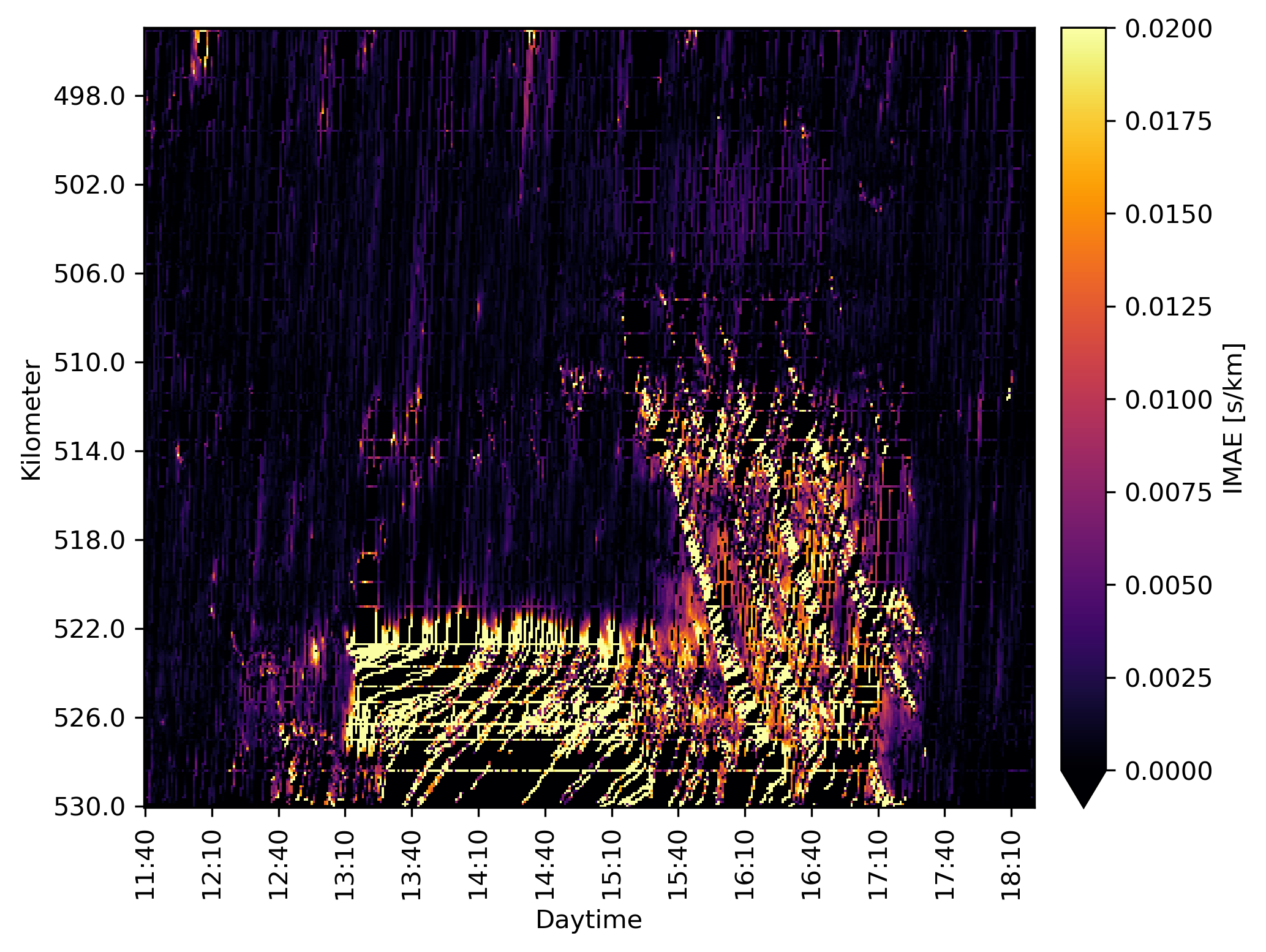}
    \end{subfigure}

	\begin{subfigure}[b]{0.41\textwidth}
		\centering
		\includegraphics[width=\textwidth]{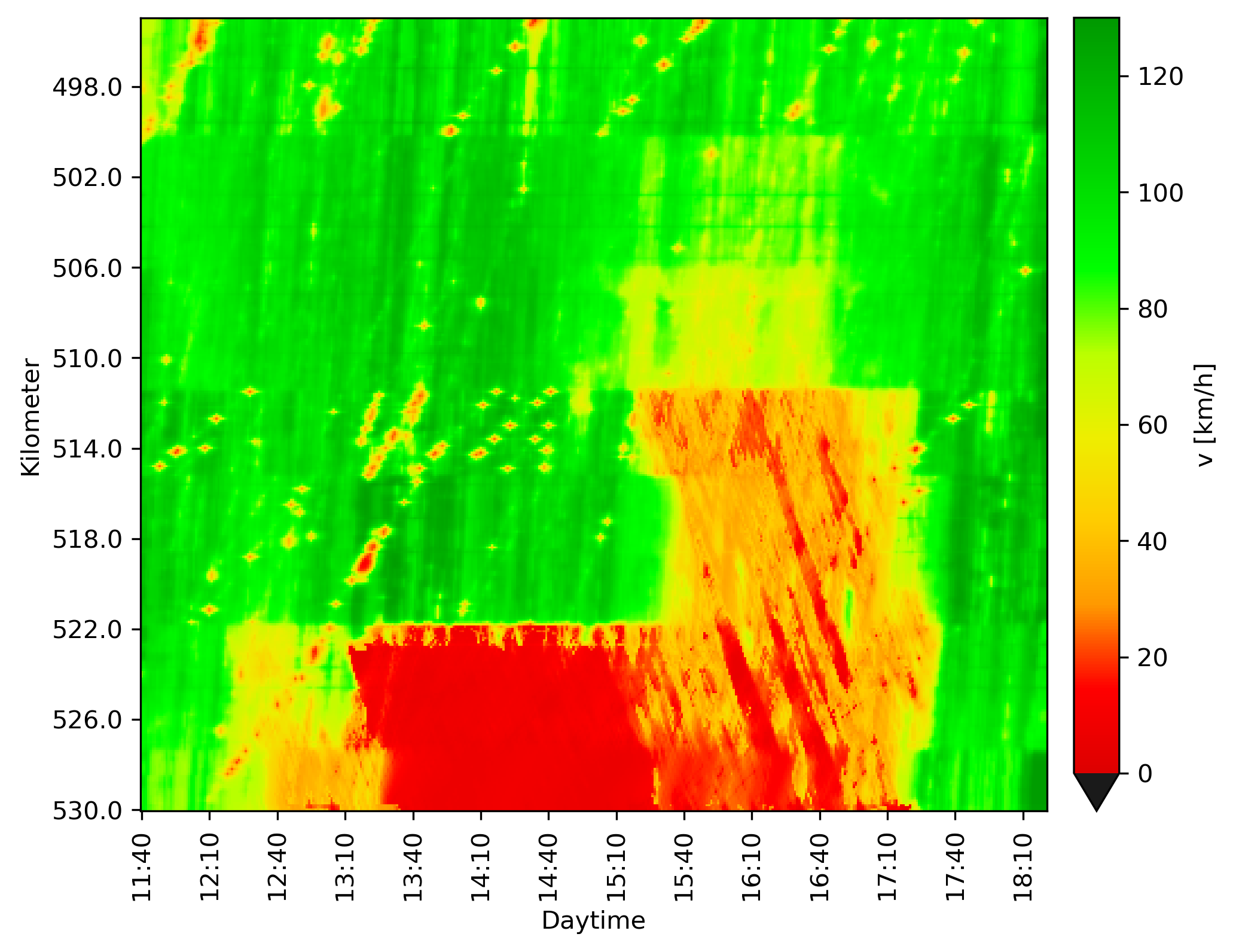}
	\end{subfigure}
    \begin{subfigure}[b]{0.41\textwidth}
    	\centering
    	\includegraphics[width=\textwidth]{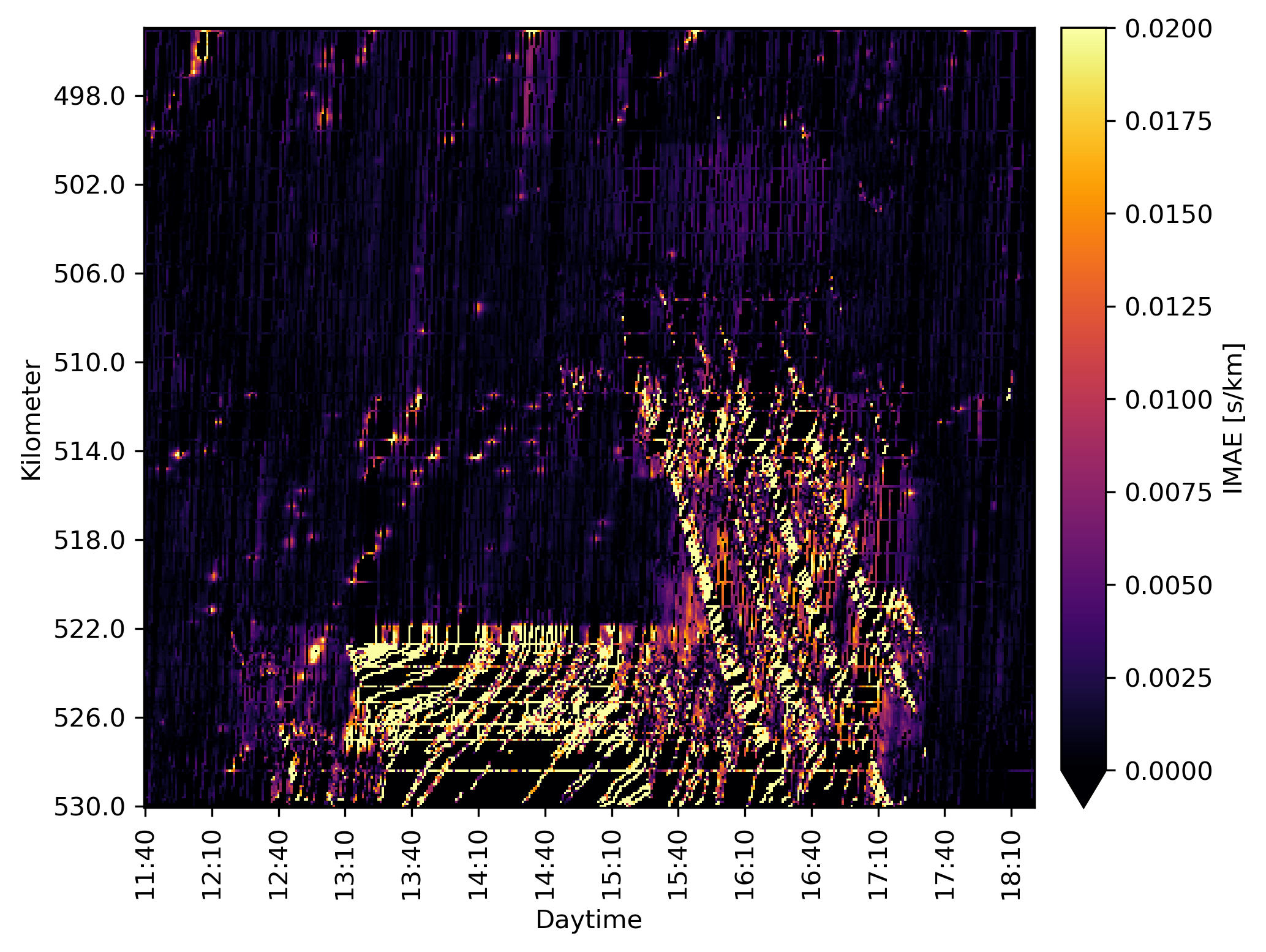}
    \end{subfigure}

	\begin{subfigure}[b]{0.41\textwidth}
		\centering
		\includegraphics[width=\textwidth]{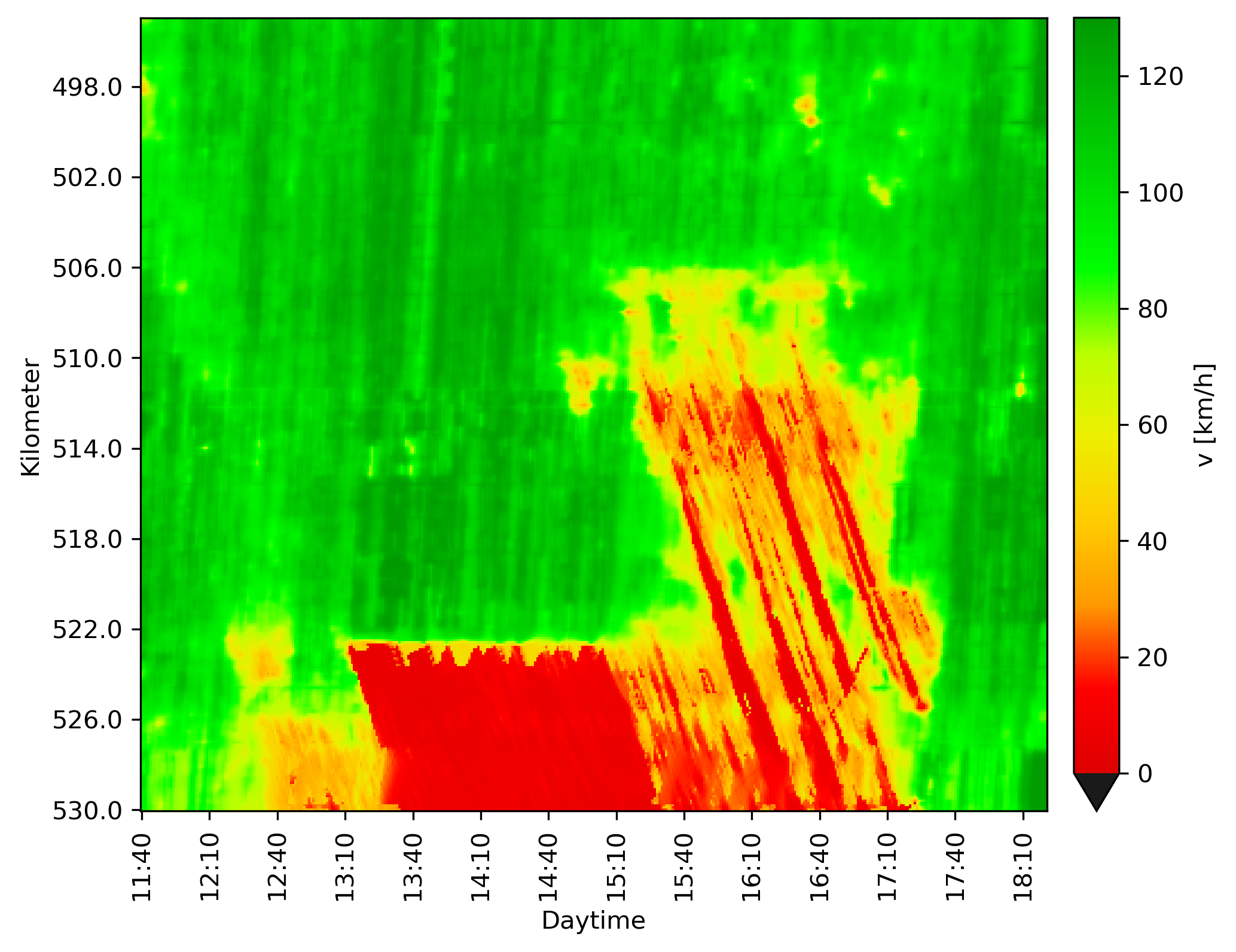}
	\end{subfigure}
	\begin{subfigure}[b]{0.41\textwidth}
		\centering
		\includegraphics[width=\textwidth]{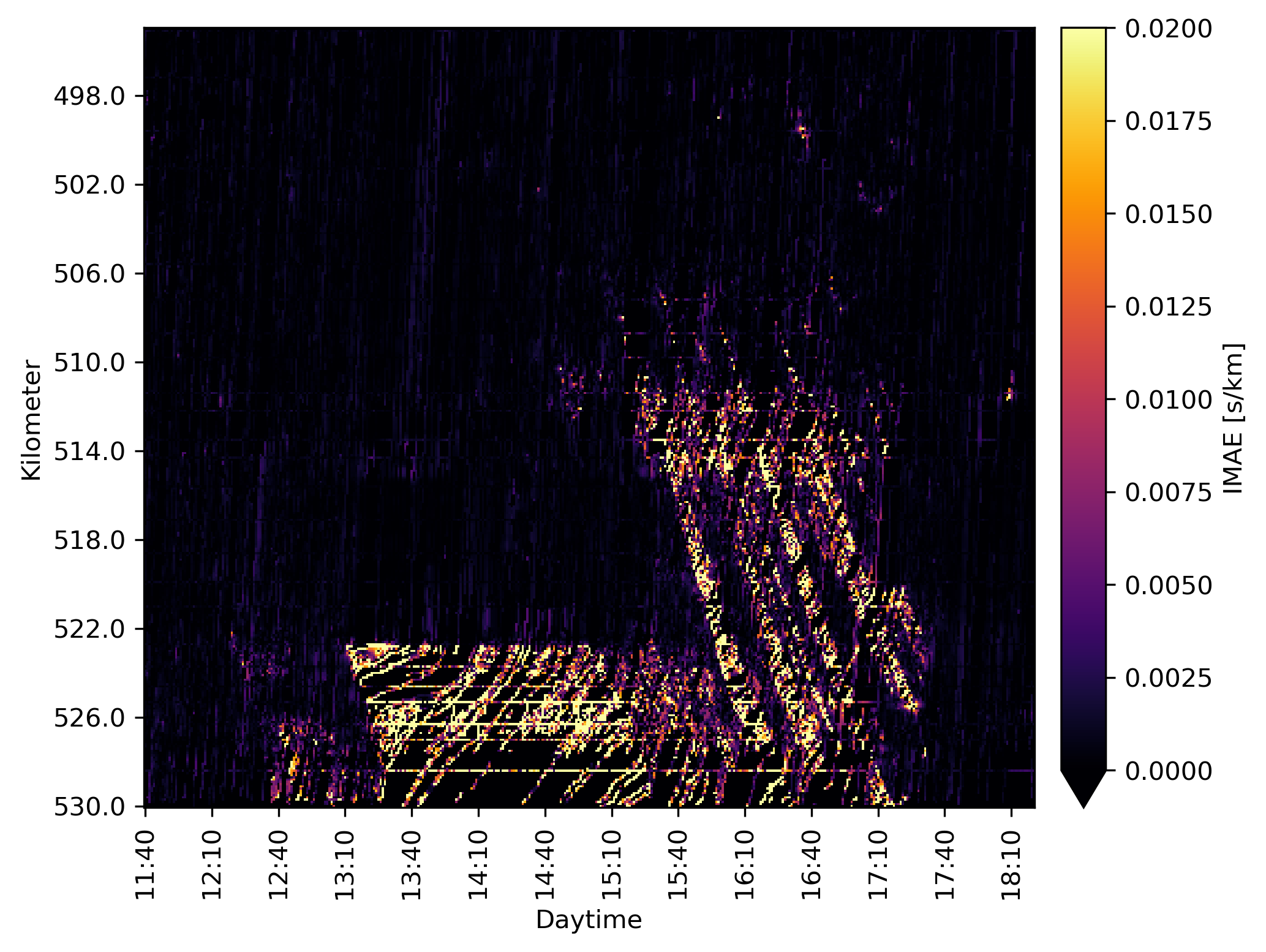}
	\end{subfigure}
    \caption{Reconstructed speeds applying each algorithm (a)~SEC-AVG, (b)~ASM, (c)~PSM, (d)~PSM-W to the training data (on the left) and resulting \acp{IMAE} comparing the reconstructed speeds to all available data (right)}
    \label{fig:recons_all}
\end{figure*}

Fig.~\ref{fig:recons_all} visualizes the estimation results of all algorithms as well as the \ac{IMAE} with respect to all available data. It can be observed that the estimate computed with the section-average approach (a) results in large errors downstream of the heavy congestion at kilometer 522. Furthermore, the approach failed to reconstruct the moving jams that emerge after 3:30pm. The reconstructions given with \ac{ASM} (b) and \ac{PSM} (c) reveal a higher spatio-temporal accuracy. Though, even these approaches spatially overestimate the heavy congestion and are not very accurate at reconstructing the moving jams either. The main reason is that all \ac{BT} data, with their low space-time accuracy in mid-range speeds (compare section \ref{sec:bt_to_v}) are smoothed, which blurs the fine structure of the congestion. 

\begin{figure}[tbh]
	\centering
	\includegraphics[width=0.7\textwidth]{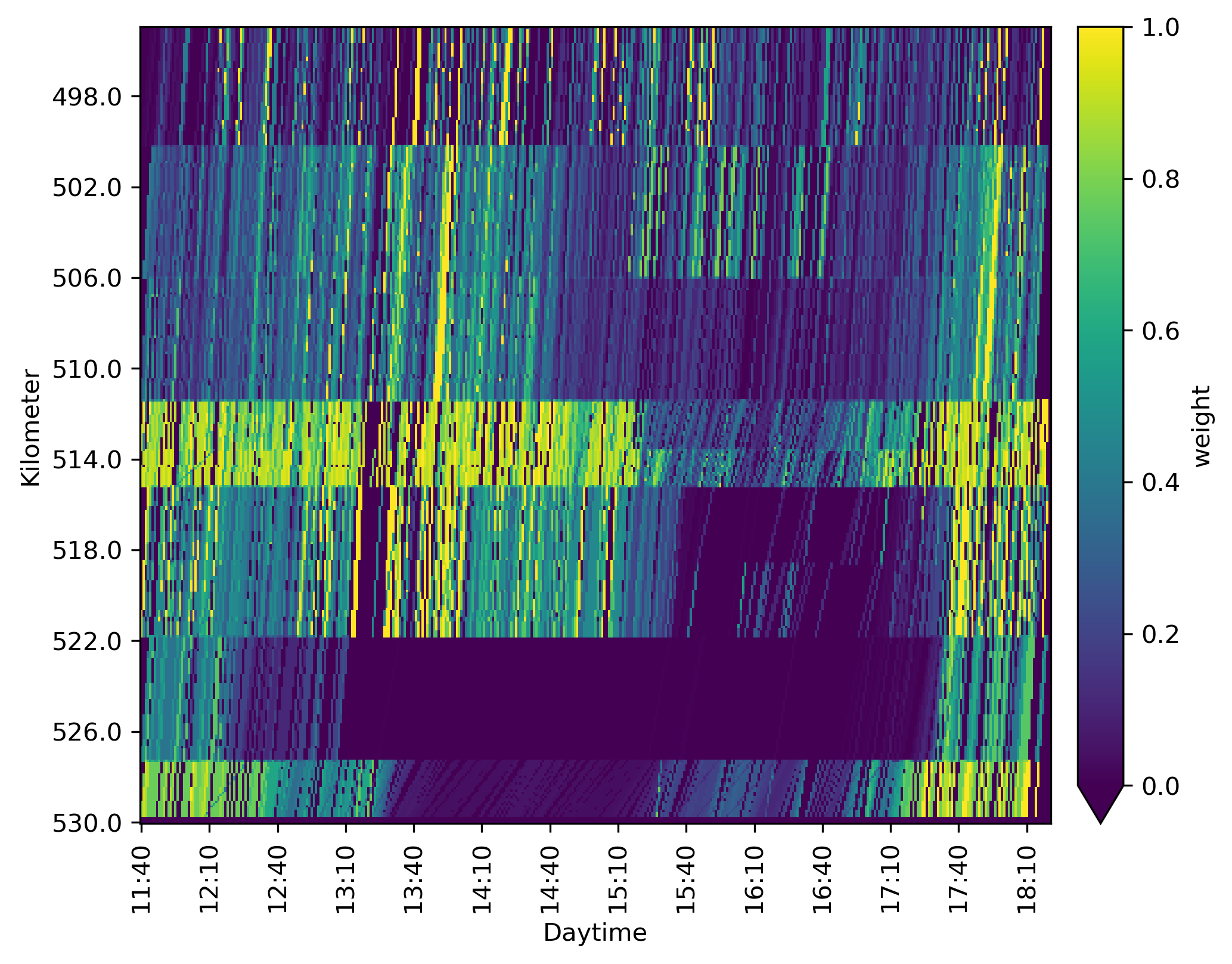}
	\caption{Resulting weight applying the speed-adaptive conversion of travel time samples}
	\label{fig:AtoP}
\end{figure}

Applying the `PSM-W' (d) with the adapted weighting of \ac{BT} according to eq.~(\ref{eq:AtoP}) (see Fig.~\ref{fig:AtoP}) overcomes this issue. Traces with medium travel times and those collected on long segments tend to have a lower weight. Thus, both the speed profile of the heavy congestion and that of the moving jams are reconstructed more precisely.

\begin{figure}[tbh]
	\centering
	\includegraphics[width=0.7\textwidth]{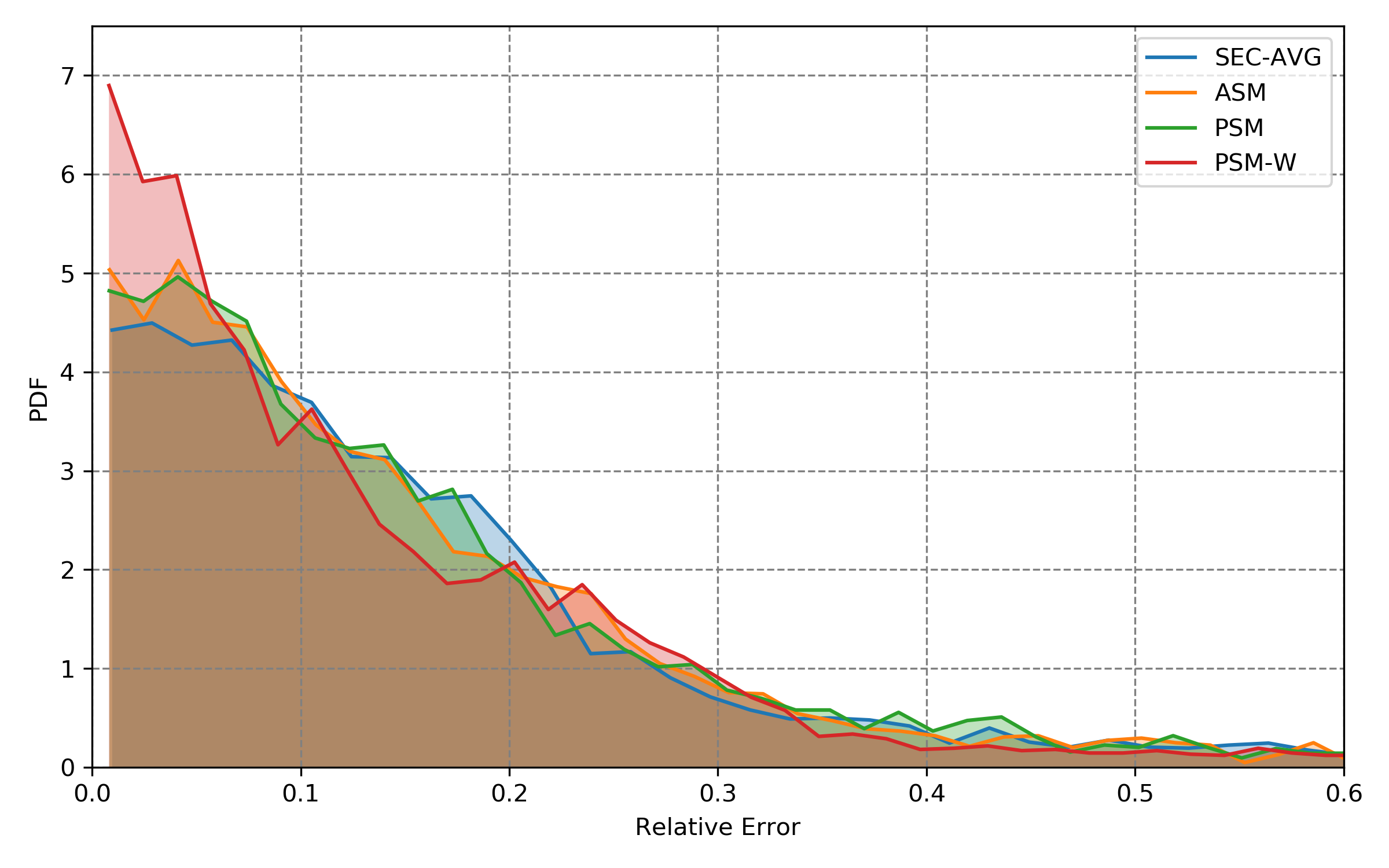}
	\caption{Approximated probability density function of relative errors comparing the travel times of virtual trajectories based on each algorithm with the measured travel times collected via \ac{BT} devices}
	\label{fig:diff_hist}
\end{figure}

Fig.~\ref{fig:diff_hist} depicts the interpolated \acp{PDF} of relative travel time errors of each algorithm based on the \ac{BT} test set and virtual trajectories (see section~\ref{sec:methodology}). The \acp{PDF} of `SEC-AVG', `ASM' and `PSM' are similar to each other, and exhibit a wider distribution than the \ac{PDF} corresponding to `PSM-W'. This explains the lower resulting \ac{MAPE} of the `PSM-W'.

\subsubsection{Sensor Setup Assessment}
\begin{figure}[tbh]
	\centering
	\includegraphics[width=0.6\textwidth]{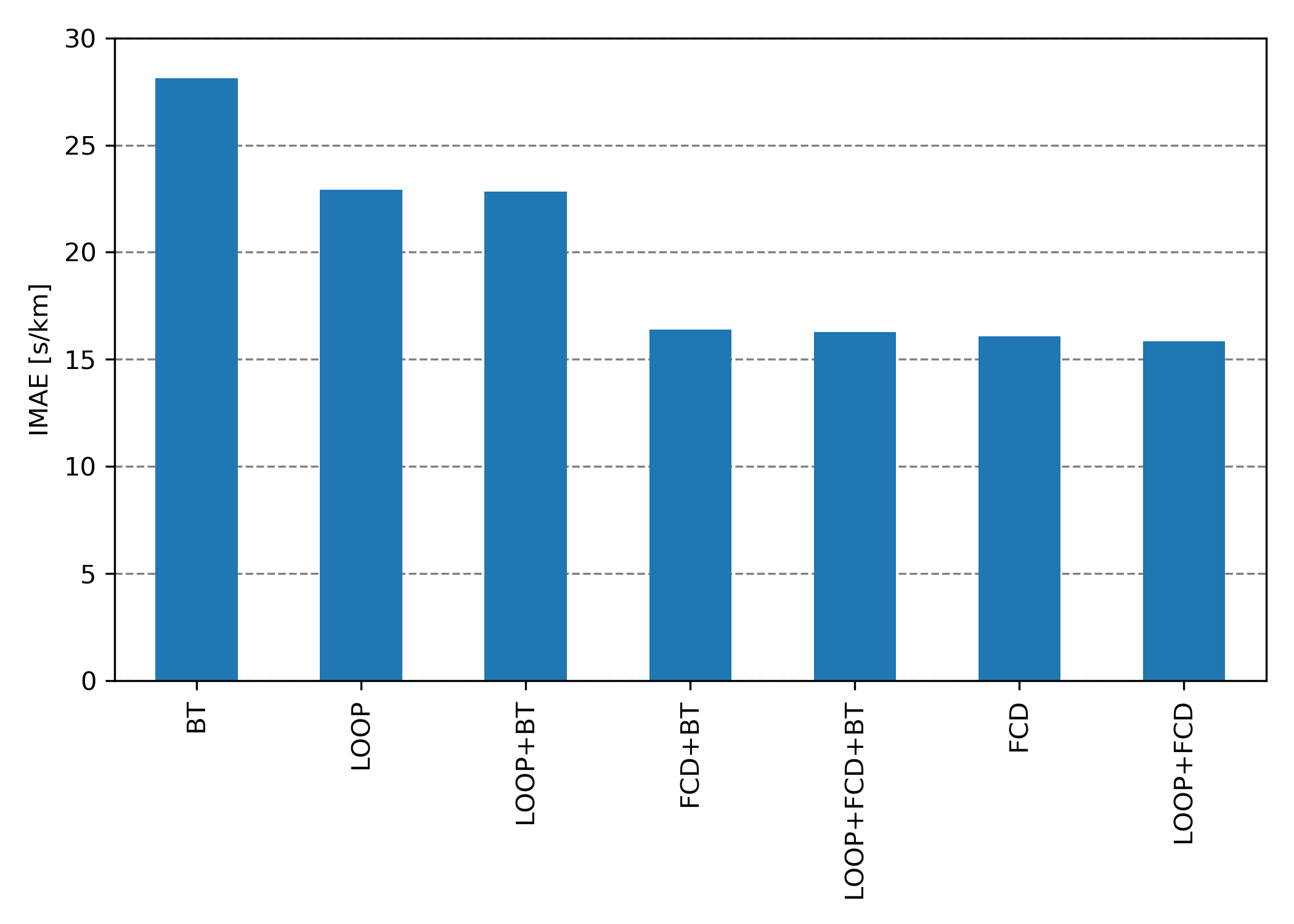} \\
	\includegraphics[width=0.6\textwidth]{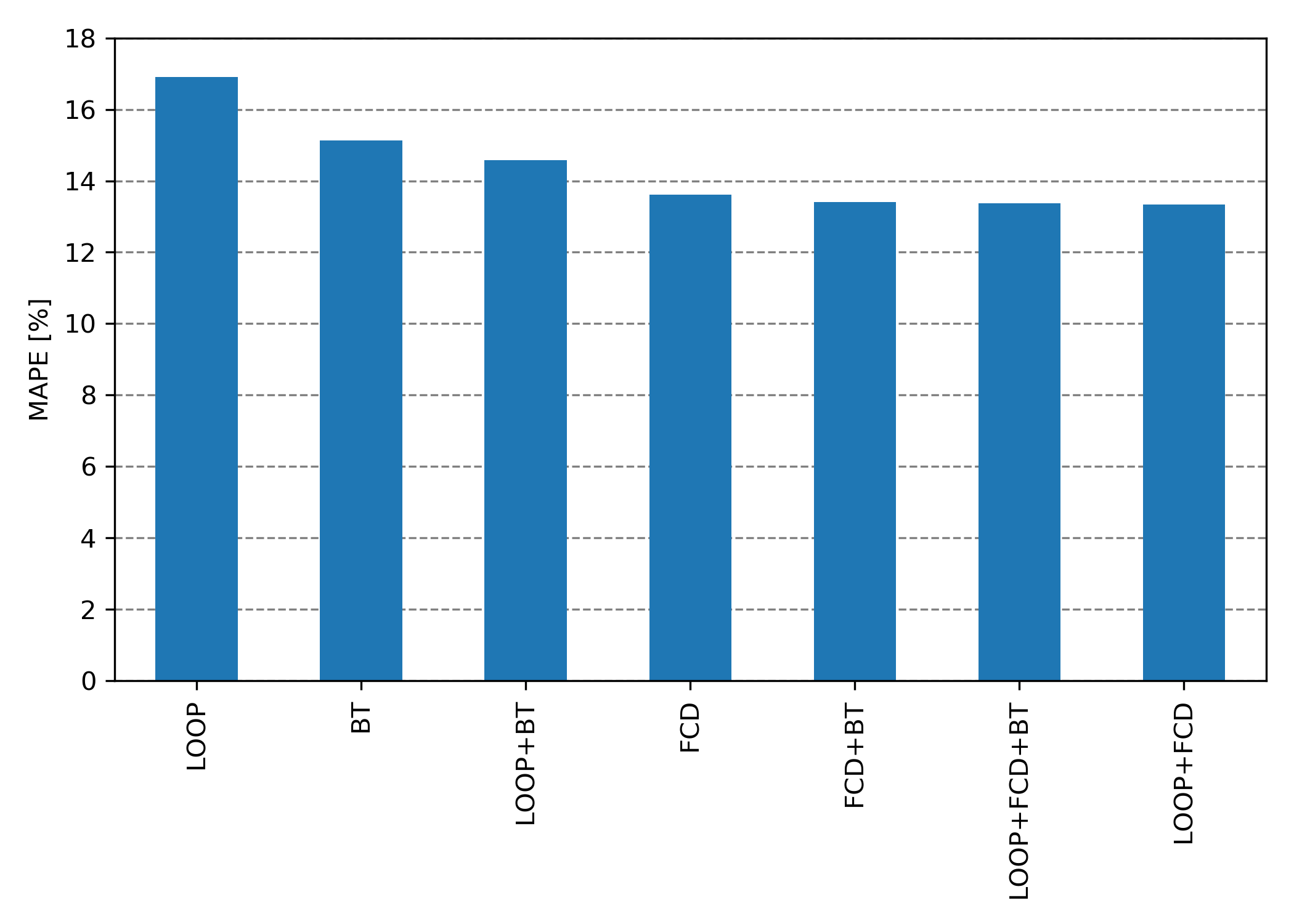} 
	\caption{Lowest \ac{IMAE} and \ac{MAPE} using the most accurate reconstruction algorithm with respect to the available data source}
	\label{fig:sensor_vs_quali}
\end{figure}

Suppose that one wishes to install an array of traffic sensors on a stretch of road for the purpose of providing accurate traffic speed information. In that case, it is relevant to know about the quality that a single sensor technology or a combination of sensor technologies may achieve. Fig.~\ref{fig:sensor_vs_quali} shows, for each sensor combination, the lowest achieved IMAE and MAPE across all algorithms. Several observations can be made:
\begin{enumerate}
	\item The combination of \ac{FCD} and loop data provides the best results for \ac{MAPE} and \ac{IMAE}.
	\item The usage of more technologies does not necessarily improve the reconstruction quality. For example, `LOOP+FCD+BT' is not the most accurate combination.
	\item With respect to \ac{IMAE}, \ac{BT} provides the lowest accuracy.
	\item With respect to \ac{MAPE}, loops provide the lowest accuracy.
	\item Using \ac{FCD} or combinations with \ac{FCD} increases both quality metrics significantly.
	\item The integration of \ac{BT} data improves the quality in some cases (\ac{MAPE}: `FCD+BT', `LOOP+BT'), but worsens it in others (\ac{IMAE}: `FCD+BT')
\end{enumerate}
Apparently, loop and \ac{FCD} is the best choice. However, if for instance \ac{FCD} are not available, a combination of loop and \ac{BT} data is able to provide more accurate results. Thus, these findings support in the decision process of setting up sensors on a road, or amending stationary data with \ac{FCD}.

\subsection{Discussion}
The present study examines two major aspects of a multi-sensor data fusion: the reconstruction accuracy using different combinations of sensor data, and the accuracy applying different state-of-the-art algorithms (as well as a novel approach) to different sensor combinations. Additionally, the reconstruction accuracy is measured using two metrics. 

A welcome result of such a study would be a clear recommendation on which algorithm or data to use in general in order to obtain the most accurate estimates. However, as the comparison showed, the choice of metric has an influence on the most accurate approach and sensor combination. For example, adding \ac{BT} data barely improved, and sometimes even worsened, the quality of the space-time speed reconstruction. On the other hand, the travel time accuracy of virtual trajectories improved by adding \ac{BT}. The same is true for the choice of algorithm. If only loop data are given, the \ac{ASM} performs best in \ac{IMAE} and \ac{MAPE}. Given other data, specially \ac{BT} data, the weighted \ac{PSM}-W performs best. Compared to the original \ac{PSM}, its accuracy is the same or better, thus, it successfully extends this approach without compromises. Thus, as a result, depending on the desired speed and travel time accuracy, this study helps to pick the optimal sensor setup or algorithm, depending on the given situation.

Some factors which may have an impact on the results are set as fixed in this study, though they may vary in other applications. First, the penetration rate and sampling interval of \ac{FCD} and the spacing of stationary detectors may vary. Secondly, the situation used for assessment in this paper is a mixture of two traffic patterns using the classification of the Three-Phase theory: mega-jam and General Pattern \cite{Kerner.1999}. These patterns cause large travel time losses, and thus, are especially important to reconstruct accurately. For further work, the study may be extended to further congestion patterns occurring on different days and roads.

\section{Conclusion}
\label{sec:conclusion}
This paper studies a multi-sensor data fusion for traffic speed and travel time reconstruction. Two aspects are analyzed: (1) Which is the most accurate algorithm depending on different combinations of data sources, and (2) which is the best performance one can achieve with a flexible sensor setup. Therefore, three state-of-the-art methods such as the \ac{ASM}, the \ac{PSM} and a simple averaging method, as well as a novel approach, are used to reconstruct the traffic speed and travel times given sparse data. 

The novel approach extends the \ac{PSM}. It introduces a variable weighting of \ac{BT} measurements, depending on detector spacing and measured travel time, which expresses the trustworthiness of a measurement. The weighting allows for a dynamic integration of \ac{BT} data with other data sources. 

The mentioned questions are studied using empirical loop data, \ac{BT} data and \ac{FCD} collected during severe congestion on a German freeway. Data are divided into a reconstruction and a test set. Various combinations of algorithms and data are used to reconstruct the space-time traffic speed and the travel times. The error metrics \ac{IMAE} and \ac{MAPE} are used to assess the resulting reconstruction accuracies.

Key findings are that the novel approach outperforms the other algorithms in most of the cases. Furthermore, a combination of \ac{FCD} and loop detector data provides the best overall results. The integration of Bluetooth data does not necessarily improve the reconstruction quality, depending on the error measure chosen. However, if no \ac{FCD} are available, a combination of loop data and \ac{BT} data is a better choice than only one source of data.

Next steps may include a mathematical optimization of the applied parameters and further studies on sensor spacings. Furthermore, the study could be extended to other locations and congestion patterns. 

\section{Acknowledgments}
The authors would like to thank \emph{Landesbaudirektion Bayern} for providing the data. This work is part of the ViM project and has been funded by the Bavarian Ministry of Economic Affairs, Regional Development and Energy (StMWi) through the Center Digitization.Bavaria, an initiative of the Bavarian State Government.

\bibliographystyle{IEEEtran}
\bibliography{arxiv_Kessler_2021}

\end{document}